\documentclass[preprint,12pt]{elsarticle}

\usepackage{amssymb}
\usepackage{comment}

\usepackage{xcolor} 

\usepackage{lscape} 
\usepackage{multicol}
\usepackage{multirow}

\newcommand{\gdso}{\mathrm{{Gd_2(SO_4)_3\cdot 8H_2O}}}
\newcommand{\GdSOw}{$\rm Gd_2(\rm SO_4)_3\cdot \rm 8H_2O$\ }
\newcommand{\GdSO}{$\rm Gd_2(\rm SO_4)_3$\ }

\journal{Nuclear Instruments and Method in Physics Research A}

\begin{document}

\begin{frontmatter}



\title{Second gadolinium loading to Super-Kamiokande}




\address[AFFicrr]{Kamioka Observatory, Institute for Cosmic Ray Research, Univerity of Tokyo, Kamioka, Gifu 506-1205, Japan}
\address[AFFkashiwa]{Research Center for Cosmic Neutrinos, Institute for Cosmic Ray Research, University of Tokyo, Kashiwa, Chiba 277-8582, Japan}
\address[AFFmad]{Department of Theoretical Physics, University Autonoma Madrid, 28049 Madrid, Spain}
\address[AFFbu]{Department of Physics, Boston University, Boston, MA 02215, USA}
\address[AFFbcit]{Department of Physics, British Columbia Institute of Technology, Burnaby, BC, V5G 3H2, Canada }
\address[AFFuci]{Department of Physics and Astronomy, University of California, Irvine, Irvine, CA 92697-4575, USA}
\address[AFFcsu]{Department of Physics, California State University, Dominguez Hills, Carson, CA 90747, USA}
\address[AFFcnm]{Institute for Universe and Elementary Particles, Chonnam National University, Gwangju 61186, Korea}
\address[AFFduke]{Department of Physics, Duke University, Durham NC 27708, USA}
\address[AFFllr]{Ecole Polytechnique, IN2P3-CNRS, Laboratoire Leprince-Ringuet, F-91120 Palaiseau, France}
\address[AFFgifu]{Department of Physics, Gifu University, Gifu, Gifu 501-1193, Japan}
\address[AFFgist]{GIST College, Gwangju Institute of Science and Technology, Gwangju 500-712, Korea}
\address[AFFglasgow]{School of Physics and Astronomy, University of Glasgow, Glasgow, Scotland, G12 8QQ, United Kingdom}
\address[AFFuh]{Department of Physics and Astronomy, University of Hawaii, Honolulu, HI 96822, USA}
\address[AFFibs]{Center for Underground Physics, Institute for Basic Science (IBS), Daejeon, 34126, Korea}
\address[AFFicise]{Institute For Interdisciplinary Research in Science and Education, ICISE, Quy Nhon, 55121, Vietnam }
\address[AFFicl]{Department of Physics, Imperial College London , London, SW7 2AZ, United Kingdom}
\address[AFFbari]{Dipartimento Interuniversitario di Fisica, INFN Sezione di Bari and Universit\`a e Politecnico di Bari, I-70125, Bari, Italy}
\address[AFFnapoli]{Dipartimento di Fisica, INFN Sezione di Napoli and Universit\`a di Napoli, I-80126, Napoli, Italy}
\address[AFFpadova]{Dipartimento di Fisica, INFN Sezione di Padova and Universit\`a di Padova, I-35131, Padova, Italy}
\address[AFFroma]{INFN Sezione di Roma and Universit\`a di Roma ``La Sapienza'', I-00185, Roma, Italy}
\address[AFFilance]{ILANCE, CNRS - University of Tokyo International Research Laboratory, Kashiwa, Chiba 277-8582, Japan}
\address[AFFkeio]{Department of Physics, Keio University, Yokohama, Kanagawa, 223-8522, Japan}
\address[AFFkek]{High Energy Accelerator Research Organization (KEK), Tsukuba, Ibaraki 305-0801, Japan}
\address[AFFkcl]{Department of Physics, King's College London, London, WC2R 2LS, UK }
\address[AFFkobe]{Department of Physics, Kobe University, Kobe, Hyogo 657-8501, Japan}
\address[AFFkyoto]{Department of Physics, Kyoto University, Kyoto, Kyoto 606-8502, Japan}
\address[AFFliv]{Department of Physics, University of Liverpool, Liverpool, L69 7ZE, United Kingdom}
\address[AFFminn]{School of Physics and Astronomy, University of Minnesota, Minneapolis, MN  55455, USA}
\address[AFFmiyagi]{Department of Physics, Miyagi University of Education, Sendai, Miyagi 980-0845, Japan}
\address[AFFnagoya]{Institute for Space-Earth Environmental Research, Nagoya University, Nagoya, Aichi 464-8602, Japan}
\address[AFFkmi]{Kobayashi-Maskawa Institute for the Origin of Particles and the Universe, Nagoya University, Nagoya, Aichi 464-8602, Japan}
\address[AFFpol]{National Centre For Nuclear Research, 02-093 Warsaw, Poland}
\address[AFFsuny]{Department of Physics and Astronomy, State University of New York at Stony Brook, NY 11794-3800, USA}
\address[AFFokayama]{Department of Physics, Okayama University, Okayama, Okayama 700-8530, Japan}
\address[AFFoecu]{Media Communication Center, Osaka Electro-Communication University, Neyagawa, Osaka, 572-8530, Japan}
\address[AFFox]{Department of Physics, Oxford University, Oxford, OX1 3PU, United Kingdom}
\address[AFFral]{Rutherford Appleton Laboratory, Harwell, Oxford, OX11 0QX, UK }
\address[AFFseoul]{Department of Physics, Seoul National University, Seoul 151-742, Korea}
\address[AFFsheff]{Department of Physics and Astronomy, University of Sheffield, S3 7RH, Sheffield, United Kingdom}
\address[AFFshizuokasc]{Department of Informatics in Social Welfare, Shizuoka University of Welfare, Yaizu, Shizuoka, 425-8611, Japan}
\address[AFFsilesia]{August Che\l{}kowski Institute of Physics, University of Silesia in Katowice, 75 Pu\l{}ku Piechoty 1, 41-500 Chorz\'{o}w, Poland}
\address[AFFstfc]{STFC, Rutherford Appleton Laboratory, Harwell Oxford, and Daresbury Laboratory, Warrington, OX11 0QX, United Kingdom}
\address[AFFskk]{Department of Physics, Sungkyunkwan University, Suwon 440-746, Korea}
\address[AFFtohoku]{Department of Physics, Tohoku University, Aoba, Sendai 9808578, Japan}
\address[AFFtokai]{Department of Physics, Tokai University, Hiratsuka, Kanagawa 259-1292, Japan}
\address[AFFtodai]{Department of Physics, University of Tokyo, Bunkyo, Tokyo 113-0033, Japan}
\address[AFFipmu]{Kavli Institute for the Physics and Mathematics of the Universe (WPI), The University of Tokyo Institutes for Advanced Study, University of Tokyo, Kashiwa, Chiba 277-8583, Japan}
\address[AFFtit]{Department of Physics,Tokyo Institute of Technology, Meguro, Tokyo 152-8551, Japan}
\address[AFFtus]{Department of Physics, Faculty of Science and Technology, Tokyo University of Science, Noda, Chiba 278-8510, Japan}
\address[AFFtriumf]{TRIUMF, 4004 Wesbrook Mall, Vancouver, BC, V6T2A3, Canada }
\address[AFFtsinghua]{Department of Engineering Physics, Tsinghua University, Beijing, 100084, China}
\address[AFFwu]{Faculty of Physics, University of Warsaw, Warsaw, 02-093, Poland}
\address[AFFwarwick]{Department of Physics, University of Warwick, Coventry, CV4 7AL, UK }
\address[AFFwinnipeg]{Department of Physics, University of Winnipeg, MB R3J 3L8, Canada }
\address[AFFynu]{Department of Physics, Yokohama National University, Yokohama, Kanagawa, 240-8501, Japan}


\author[AFFicrr,AFFipmu]{K.~Abe}
\author[AFFicrr]{C.~Bronner}
\author[AFFicrr,AFFipmu]{Y.~Hayato}
\author[AFFicrr,AFFipmu]{K.~Hiraide}
\author[AFFicrr]{K.~Hosokawa}
\author[AFFicrr,AFFipmu]{K.~Ieki}
\author[AFFicrr,AFFipmu]{M.~Ikeda}
\author[AFFicrr,AFFipmu]{J.~Kameda}
\author[AFFicrr]{Y.~Kanemura}
\author[AFFicrr]{R.~Kaneshima}
% ...

\begin{abstract}

The first loading of gadolinium (Gd) into Super-Kamiokande in 2020 was successful, and the neutron capture efficiency on Gd reached 50\%. To further increase the Gd neutron capture efficiency to 75\%, 26.1 tons of \GdSOw was additionally loaded into Super-Kamiokande (SK) from May 31 to July 4, 2022. As the amount of loaded \GdSOw was doubled compared to the first loading, the capacity of the powder dissolving system was doubled. We also developed new batches of gadolinium sulfate with even further reduced radioactive impurities. In addition, a more efficient screening method was devised and implemented to evaluate these new batches of \GdSOw.
Following the second loading, the Gd concentration in SK was measured to be $333.5 \pm2.5$ ppm via an Atomic Absorption Spectrometer (AAS). From the mean neutron capture time constant of neutrons from an Am/Be calibration source, the Gd concentration was independently measured to be  332.7~$\pm$~6.8(sys.)~$\pm$~1.1(stat.) ppm,  consistent with the AAS result. Furthermore, during the loading the Gd concentration was monitored continually using the capture time constant of each spallation neutron produced by cosmic-ray muons, and the final neutron capture efficiency was shown to become 1.5 times higher than that of the first loaded phase, as expected.

\end{abstract}






\end{frontmatter}


\section{Introduction} 
\label{intro}
In 2020, a new phase of Super-Kamiokande (SK), SK-Gd, was started by loading 13 tons of \GdSOw into pure water~\cite{Gd1st,SK6SRN}. This led to a high neutron detection efficiency, which allows us to distinguish different neutrino interactions, enhance signals and remove backgrounds more efficiently~\cite{Beacom:2003nk}. In the summer of 2022, an upgrade of SK-Gd was performed by adding additional Gd. Thus a total of about 40 tons of gadolinium sulfate octahydrate has been introduced into the SK water, increasing the Gd neutron capture efficiency to 
75\%~\cite{EGADS}.

In this paper, we report the details of the second Gd loading, including the upgraded Gd dissolving system and the pre-treatment of Gd-dissolved water, the properties of the dissolved \GdSOw, as well as calibration measurements to confirm the loaded Gd concentration.
In Section 2, we describe the \GdSOw dissolving system that has been increased to twice the capacity and its operation scheme, and, in Section 3, the details of the ion exchange resin needed to treat \GdSO dissolved water. In Section 4, we describe the specification of the prepared \GdSOw and the details of the screening of 27 tons of \GdSOw. The Gd loading for this second stage of SK-Gd is described in Section 5, and the results of the measurements of Gd concentration in SK are reported in Section 6. Concluding remarks are presented in Section 7.

\section{Improved Dissolving System}
\label{sec:system}
\subsection{Overview of the water flow}
\label{sec:waterflow}
\begin{figure}[htbp]
    \centering
    \includegraphics[width=0.8\textwidth]{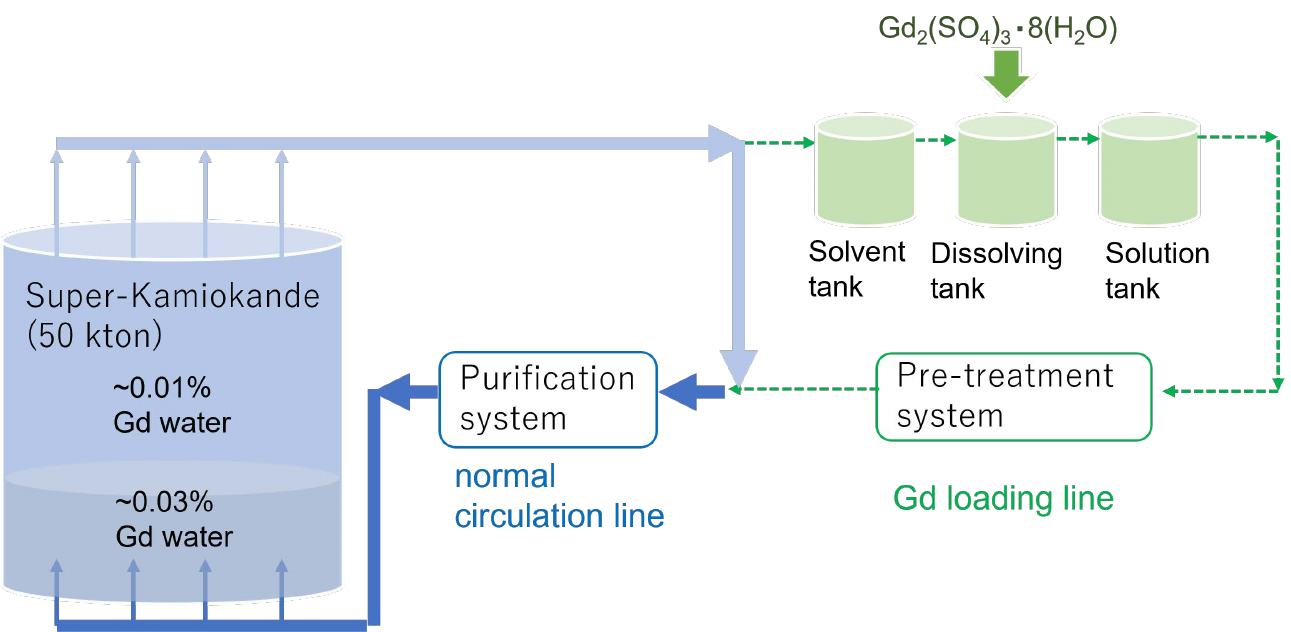}
    \caption{Overall water flow of SK during the dissolving work.}
    \label{fig:overall_system}
\end{figure}

For the second loading, the required amount of $\gdso$ powder is doubled from the previous loading. 
Therefore, improving the dissolving speed by a factor of two was required. 
Figure~\ref{fig:overall_system} shows an overview of the water flow of SK during the dissolving work.
The overall system is unchanged from the 1st loading; it is described in detail in~\cite{Gd1st}.  

During the dissolving work, the Gd loading line is connected to the normal water circulation line.
Part of the water in the normal circulation flow is diverted to the solvent tank of the 
Gd loading line and the $\gdso$ is added to the dissolving tank. 
High concentration Gd water is transported to the solution tank, and the solution tank continuously supplies the Gd water back to the normal circulation line through the pre-treatment system, which contains filters, UV sterilizers, and ion exchange resins to remove impurities in the water. 

The new Gd-loaded water is supplied from the bottom of the SK tank and returns to the circulation line
from the top. The supply water temperature was controlled to be colder than the tank water by 
$\sim0.3~^\circ$C to suppress convection. In this way, the Gd concentration of the water
supplied to the dissolving tank during this 2nd Gd loading is kept at the 1st Gd loading level of $\sim$0.01\%. 
With the 60~m$^3$/hour flow rate of the normal circulation line, it takes $\sim$1 month for
the $\sim$0.03\% Gd water to fill up the SK tank. 
Since 27.3 tons of Gd sulfate powder can be dissolved in one month, this layered approach is the most efficient way to achieve a uniform $\sim$0.03\% Gd concentration by utilizing a single turnover of the water in the SK tank.

\subsection{The dissolving system}

\begin{figure}[htbp]
    \centering
    \includegraphics[width=0.8\textwidth]{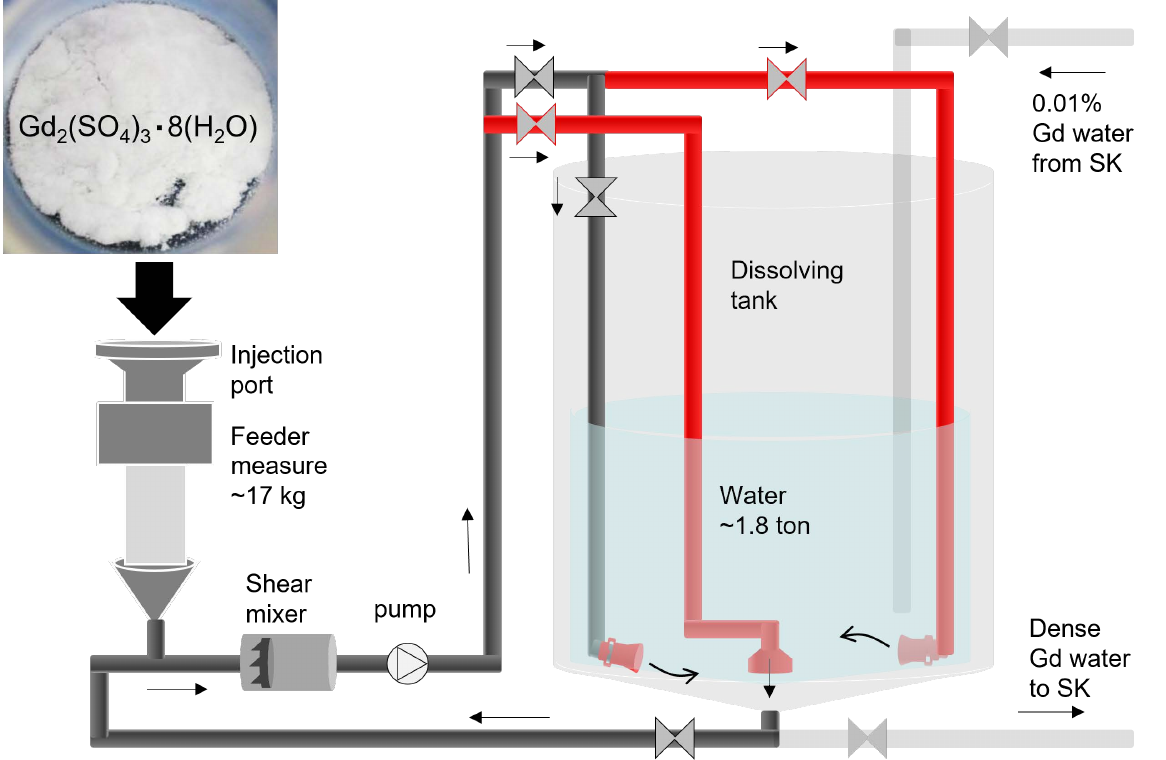}
    \caption{The dissolving system.}
    \label{fig:dissolving_system}
\end{figure}

Figure \ref{fig:dissolving_system} shows the improved dissolving system: the dissolving tank (4~m$^3$), the Gd injection port, and the shear mixer.
Dissolving was accomplished via the following procedure:
\begin{enumerate}
    \item Load $\sim$400~kg of $\gdso$ into the injection port whenever the feeder becomes almost empty. 
    This work was done manually twice per day, after which the subsequent works were 
    performed automatically.
    \item Fill the dissolving tank with $\sim$1.8~m$^3$ of water from the solvent tank.
    \item Start water circulation between the shear mixer and the dissolving tank.
    \item The feeder measures the weight of the powder so that it contains and 
    supplies $\sim$17~kg of $\gdso$ powder to the circulation line.
    \item Continue water circulation for $\sim$15 minutes until all the powder is broken apart 
    by the shear mixer and dissolved into water.
    \item After dissolving, transport high concentration Gd water to the solution tank.
\end{enumerate}

To improve the loading rate by a factor of two, we double the amount of powder and water from the previous loading. 
In addition, it was necessary to improve the pipe design for the circulation. 
There are three pipes for circulation inserted from the top of the dissolving tank, as shown in Figure \ref{fig:dissolving_system}, but 
only the leftmost one was used in the previous Gd loading. This pipe is bent 90 degrees
at the bottom of the tank to make a vortex inside the tank.
If the vortex is not strong enough, the powder tends to stagnate at the bottom of the 
dissolving tank without returning to the shear mixer. 
As we doubled the amount of water in the dissolving tank, it was more difficult to make a vortex inside the tank.
Therefore, we added two more pipes for the circulation. 
The pipe which goes directly to the exit of the tank efficiently returns
the powder to the shear mixer.
This pipe was used during the first $\sim$5 minutes of the dissolving process.
This line was then partially closed, and the other two lines with 90-degree bends
were opened to make a vortex for the remaining 15 minutes. Water ejectors were attached to 
the outlets of these pipes to make the flow faster. 
The discharge pressure of the circulation pump became too high with the ejector, 
so the second pipe with a 90-degree bend was added to the other side of the existing pipe 
to reduce the discharge pressure and to make the flow more uniform in the tank.

Furthermore, the powder injection port was enlarged to make the powder injection work easier.
A wire mesh net with a 3~cm mesh size was added at the entrance of the port to avoid
injecting clumps of powder.

\section{Resins in the pre-treatment system}
In the Gd-water purification system, any ionized impurities, including some radioactive impurities such as uranium and radium, are removed by ion-exchange resins. We developed special resins for the first Gd-loading, as explained in~\cite{Gd1st}. These resins have been modified to contain gadolinium or sulfate as the ion exchange groups such that the resin’s cation or anion exchange action never results in a loss of dissolved gadolinium sulfate content.
However, the basis resins for the cation exchange resin (AMBERJET\texttrademark 1020~\cite{AJ1020}) and the anion exchange resin (AMBERJET\texttrademark 4400~\cite{AJ4400}) have since been discontinued. Therefore, new resins were developed for the second Gd-loading. For the new foundational resins, a strongly acidic cation exchange resin AMBERLITE\texttrademark IR120B~\cite{AL120B} and a strongly basic anion exchange resin AMBERLITE\texttrademark 4002~\cite{AL4002} were selected.

Before and after the first Gd-loading, the cation exchange resin AMBERJET\texttrademark 1020(Gd) of the pre-treatment system was sampled and measured with the high-purity Ge detector~\cite{SKGe} described in Sec.~\ref{sec:Ge}.
The results are shown in Table~\ref{tab:usedresin} -- it was confirmed that $^{226}$Ra was captured by AMBERJET\texttrademark 1020(Gd).
The remaining ion exchange capacity of the resin was also checked after the first Gd-loading, and it was determined that the resin itself was not broken during the 1st loading; there was still removal capacity available afterward. But since new radio isotopes (RI) such as $^{222}$Rn could be expected to be emitted as the adsorbed $^{222}$Ra decayed, this partially-used cation resin was replaced with AMBERLITE\texttrademark IR120B(Gd) before the second Gd-loading.
The anion exchange resin AMBERJET\texttrademark 4400(SO$_4$) of the pre-treatment system was also evaluated after the first loading.
We could not identify any degradation of the resin nor any risk related to retaining it; therefore, the new AMBERLITE\texttrademark 4002(SO$_4$) was not deployed, and the original AMBERJET\texttrademark 4400(SO$_4$) was also used for the second loading.

\begin{table}[!ht]
    \centering
    \begin{tabular}{llc} \hline
                          &           & $^{226}$Ra concentration \\
    period                & sample    & in resin(mBq/kg) \\ \hline \hline
    1st loading in 2020 & AJ1020(Gd) before using & 1.28$\pm$0.24 \\
                          & AJ1020(Gd) after using  & 4.90$\pm$1.59 \\ \hline
    2nd loading in 2022 & AL IR120B(Gd) before using  & $<$0.99 \\
                          & AL IR120B(Gd) after using   & 1.10$\pm$0.30 \\ \hline
    \end{tabular}
    \caption{\label{tab:usedresin}
    HPGe measurement results of resins before and after using in the 1st and 2nd  $\gdso$ loading of 13 and 27 tons, respectively.
    }
\end{table}


\section{Gd sulfate powder for the second loading}

\subsection{Required quality and amount of ultra-pure Gd sulfate powder}
The \GdSOw for SK-Gd had to fulfill the requirements for impurities described in~\cite{1stGdProdScrn}. The requirements and measured values for the summed average of batches of \GdSOw are shown in  Table~\ref{tab:RIsummary}.

The criteria are set such that the additional event rate due to radioactive impurities in the powder is lower than the unloaded background rate of solar neutrinos or diffuse supernova neutrino background (DSNB) searches in SK, even after loading to the final target concentration of 0.1\% Gd. 
As shown in \cite{1stGdProdScrn}, backgrounds for the DSNB are estimated from the fraction of the spontaneous fissions of $^{238}$U yielding one neutron and one $\gamma$, where the gamma has a reconstructed energy of 10-20 MeV based upon the $\gamma$ spectrum measured in \cite{SFspec}.

In order to meet the requirements, chemical processing procedures were developed by an extensive R\&D program, which is also explained in~\cite{1stGdProdScrn}.
During the second loading, 27.3 tons of powder was dissolved into the detector. 
In addition to the chemically-bound octahydrate, 
the powder contained residual water left over from processing which averaged 4.5\%. Therefore, the mass of \GdSOw itself is 26.1 tons.
This amount (plus Gd from the first loading) should yield a Gd concentration in the SK tank of 0.033\%, equivalent to an anhydrous gadolinium sulfate ($\rm Gd_2(\rm SO_4)_3$) concentration of 0.079\%. 

\subsection{ICP-MS for U, Th and Ce}
As we did in the first loading, Inductively Coupled Plasma Mass Spectrometry (ICP-MS) is used to assay U, Th, and Ce impurities prior to high-purity germanium (HPGe) gamma spectrometries. 

To measure U and Th at the parts-per-trillion (ppt) level, 
we adopted the screening method used in the first loading~\cite{1stGdProdScrn}.
In the method, to separate U and Th from the Gd, a nitric acid aqueous solution in which a sample of \GdSOw\ is dissolved is first passed through a well-washed chromatographic extraction resin, which absorbs about 90\% or more U and Th. Then, U and Th can be eluted from the resin with a dilute nitric acid solution. Finally, the trace amounts of U and Th in the eluate can be measured by ICP-MS without interference from the Gd, which has been reduced by a factor of about 10$^4$ by this process~\cite{UTEVA}.

To assay Ce impurities, \GdSOw\ is diluted with 2\% HNO$_3$
to a mass ratio of 1 part in 10,000.
Since the matrix effect due to the existence of Gd is negligible at this concentration (0.01\%),
the concentration of Ce in the solution can be directly measured using ICP-MS. 

 It was confirmed that all samples used in the second loading meet the criteria for U, Th, and Ce contamination. The detailed results are shown in~\ref{apdx:ICP-MSres}.

\subsection{Radioisotope measurement using HPGe detectors}
\label{sec:Ge}
As was done during the first $\gdso$ loading, HPGe detectors are again used to measure the concentration of long-lived nuclides based on the gamma-ray emission of the parent or its progeny~\cite{1stGdProdScrn}.
The HPGe detectors are located at the Boulby UnderGround Screening (BUGS) facility in northern England, the {\it Laboratorio Subterráneo de Canfranc} (LSC) in Spain, and the Kamioka Observatory in Japan~\cite{SKGe}.
HPGe $\gamma$ spectrometry evaluates the concentration of $\rm{^{238}U}$, $\rm{^{226}Ra}$, $\rm{^{228}Ra}$, $\rm{^{228}Th}$, $\rm{^{235}U}$, and $\rm{^{227}Ac}$ decay series.
In addition, concentrations of $^{40}$K, $^{138}$La, $^{176}$Lu, $^{134}$Cs, and $^{137}$Cs were measured.
The various RI contaminations in every lot were screened by HPGe detectors prior to dissolution, except for the $\rm{^{226}Ra}$ in the last two lots delivered.
Most batches were screened by HPGe detectors at more than one laboratory to assess consistency in measurements.
The detailed results are shown in \ref{app:Ge}.

In order to wait for the decay of background $\rm{^{222}Rn}$, an  HPGe measurement for a $\gdso$ sample in Kamioka takes approximately 20~days.
The last two lots, 220691 and 220603, were delivered to the SK-Gd site about one week before their dissolution.
There was not enough time to measure $\rm{^{226}Ra}$ contamination after waiting for the decay of background $\rm{^{222}Rn}$.
Therefore, a chemical separation method was applied to the two lots to evaluate the $\rm{^{226}Ra}$ concentration~\cite{RaPaper}.
The $^{226}$Ra concentrations in the $\gdso$ samples measured by the chemical separation method are 0.84$\pm$0.05 and 0.23$\pm$0.02 mBq/kg.
As these concentrations include the procedure blank, the intrinsic concentrations would be lower than these values.
Consequently, we determined that the $^{226}$Ra concentrations levels in these two samples are acceptable.
The HPGe results for the $\rm{^{226}Ra}$ concentration in the two lots measured after their dissolution can be found in \ref{app:Ge}.

\subsection{RI summary for physics}
Table~\ref{tab:RIsummary} summarizes the HPGe and ICP-MS measurement results for all the batches of $\gdso$ for the second Gd loading, and the combined results for both the first and the second Gd loadings.
The requirements for each RI decay chain series are also shown.
The \textit{total budget} describes the acceptable decay rate in SK-Gd assuming that 130 tons of $\gdso$ will eventually be loaded (0.1\% Gd concentration).
The \textit{finite value} in the HPGe column shows the sum of all finite measured activities, with the errors combined in quadrature.
The \textit{upper limit} shows a conservative upper bound on the total activity, the sum of all 95\% confidence level upper limits.

The ICP-MS results indicate that $^{232}$Th and $^{238}$U contamination are sufficiently low.
Loading of more uniform quality and cleaner $\gdso$ than the 1st loading work were achieved in this loading operation.

The later parts of the chains were measured using HPGe detectors.
The activities of the $^{238}$U decay chain ($^{238}$U and $^{226}$Ra equivalent) are sufficiently small, even when the first loading contamination is added.
The $^{235}$U decay chain contamination ($^{235}$U and $^{227}$Ac equivalent) is measured only by HPGe detectors.
The resulting 95\% upper limits for the decay chain are also sufficiently small compared to the total budget.
The upper limits for the $^{232}$Th decay chain($^{228}$Ra and $^{228}$Th equivalent) are comparable to the 6.5~Bq budget.
Since we do not have good enough sensitivity for the $^{232}$Th decay chain, it is difficult to confirm whether the 6.5~Bq budget is achieved.
In addition, the 95\% upper limits for the $^{232}$Th decay chain were $<$11~Bq for $^{228}$Ra and $<$8.9~Bq for $^{228}$Th in the first loading.
Since their half-lives are 5.75~y and 1.9~y and they decay for 2~years from 2020 to 2022, the upper limits for the activities of $^{228}$Ra and $^{228}$Th are $<$8.65~Bq and $<$9.3~Bq, respectively.
The upper limit of $^{228}$Th has increased due to decays of $^{228}$Ra.
The total activities of $^{228}$Ra and $^{228}$Th after adding the contamination in the first Gd loading are $<$18.6~Bq and $<$15.9~Bq, respectively.

\begin{table}[!ht]
    \centering
    \caption{Summary of the HPGe and ICP-MS measurement results for the second Gd-loading compared with the total SK-Gd radioactivity budget assuming 0.1\% Gd-loading (130 tons of $\gdso$).
    The measurements of each radioactive chain are separated into those for the parent radioactive isotopes (RIs), the early part of the chain (E), and the late part of the chain (L).
    The HPGe assay results are combined in two ways to give an estimate of the minimum and maximum total added radioactivity to SK.
    }
    \scalebox{0.63}[0.63]{
    \begin{tabular}{ccccccccccccccccc} \hline
         \multirow{5}{*}{Chain}       &    & &                               & & \multicolumn{4}{c}{This work}              & & \multicolumn{4}{c}{Total to date} \\ \cline{6-9} \cline{11-14}
                                      &    & \multicolumn{2}{c}{Requirement} & & \multicolumn{2}{c}{HPGe} & & ICP-MS        & & \multicolumn{2}{c}{HPGe}  & & ICP-MS \\ \cline{3-4} \cline{6-7} \cline{9-9} \cline{11-12} \cline{14-14}
                                      & Part of          & Specific & Total  & & Finite        & Upper    & & Total         & & Finite        & Upper    & & Total \\
                                      & Chain            & Activity & Budget & & Value         & Limit    & & (Bq)          & & Value         & Limit    & & (Bq)  \\
                                      &                  & (mBq/kg) & (Bq)   & & (Bq)          & (Bq)     & &               & & (Bq)          & (Bq)     & &       \\ \hline \hline
          \multirow{3}{*}{$^{238}$U}  & RI $^{238}$U     & $<$5     & 650    & & --            & --       & & 0.54$\pm$0.01 & & --            & --       & &  0.88$\pm$0.15   \\
                                      & E, $^{238}$U Eq. & $<$5     & 650    & & 0             & $<$183   & & --            & & 0             & $<$272   & &  --   \\
                                      & L,$^{226}$Ra Eq. & $<$0.5   & 65     & & 3.76$\pm$0.43 & $<$10    & & --            & & 4.0$\pm$0.4   & $<$15.6  & &  --   \\
          \hline
          \multirow{3}{*}{$^{232}$Th} & RI $^{232}$Th    & $<$0.05  & 6.5    & & --            & --       & & 0.21$\pm$0.01 & & --            & --       & &  0.46$\pm$0.07   \\
                                      & E, $^{228}$Ra Eq.& $<$0.05  & 6.5    & & 2.14$\pm$0.48 & $<$11    & & --            & & 5.4$\pm$0.6   & $<$19.7  & &  --   \\
                                      & L, $^{228}$Th Eq.& $<$0.05  & 6.5    & & 1.89$\pm$0.4  & $<$8     & & --            & & 5.6$\pm$0.5   & $<$17    & &  --   \\
          \hline
          \multirow{2}{*}{$^{235}$U}  & E, $^{235}$U Eq. & $<$30    & 3900   & & 0             & $<$22    & & --            & & 4.1$\pm$0.8   & $<$37    & &  --   \\
                                      & L, $^{227}$Ac Eq.& $<$30    & 3900   & & 0             & $<$23    & & --            & & 3.3$\pm$0.7   & $<$42    & &  --   \\ 
    \hline
    \end{tabular}
    }
    \label{tab:RIsummary}
\end{table}


\section{The second gadolinium loading of Super-Kamiokande}

\subsection{Water flow and temperature}
\begin{figure}[htbp]
    \centering
    \includegraphics[width=1.0\textwidth]{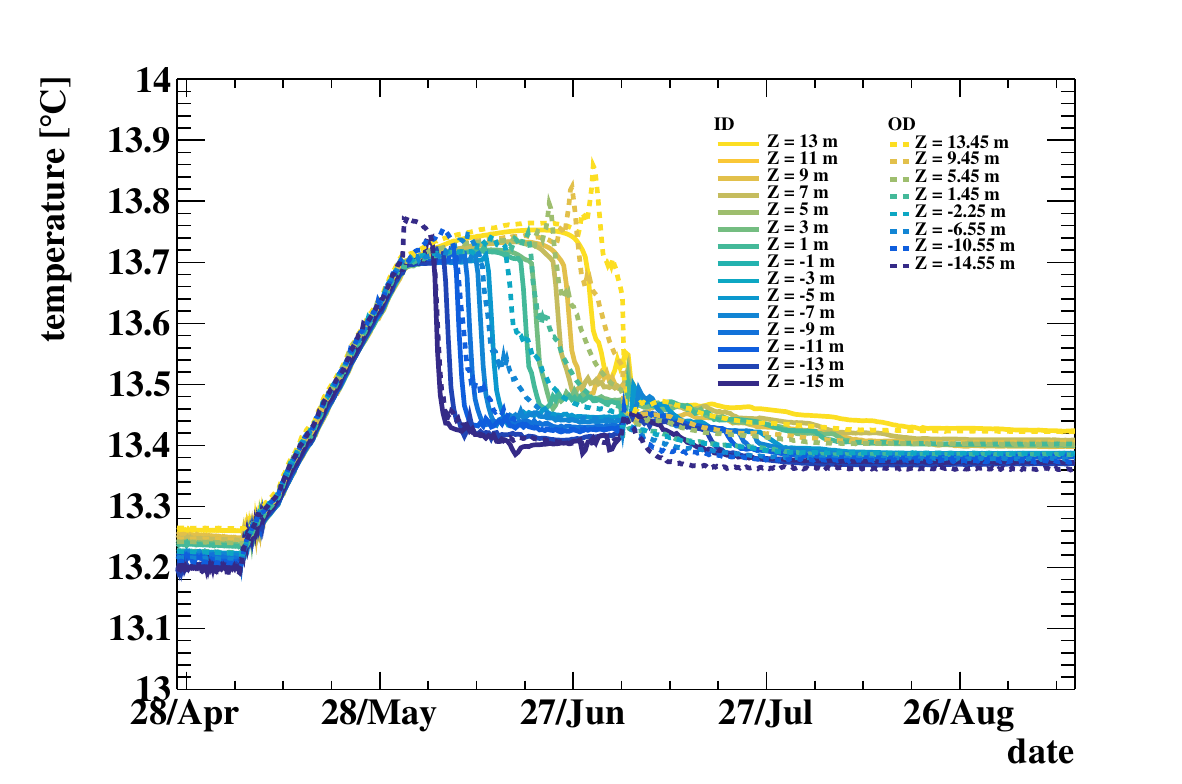}%
    \caption{Temperature of the SK tank water at different $z$ positions in the tank. $z=0$ corresponds to the center of the tank, while positive(negative) values of $z$ are in its upper(lower) half.\label{fig:temperature}}
\end{figure}

As described in Sec.~\ref{sec:waterflow}, during normal data-taking operations the temperature of the supply water must be lower than the SK tank water to avoid evoking convection. 
Before the second Gd-loading, from May 6th to May 30th, 2022, the SK supply water temperature was gradually raised from 13.65 $^\circ$C to 14.10 $^\circ$C to raise the temperature in the tank. Figure~\ref{fig:temperature} shows the temperature of the water inside the SK tank. The SK tank is separated into ID (Inner Detector) and OD (Outer Detector) regions, with temperature sensors located at different heights in each region. When the supply water temperature was raised, the temperature became uniform throughout the tank volume due to convection. Then we started to supply water with high Gd concentration ($\sim0.03\%$) and low temperature. As shown in Fig.~\ref{fig:temperature}, the cold water region, which corresponds to the high Gd concentration region, gradually extended from the bottom to the top. The supply water flow balance between ID and OD was controlled so that the height of the cold water front became the same in both regions.

\subsection{Working record}
\begin{figure}[htbp]
    \centering
    \includegraphics[width=1.0\textwidth]{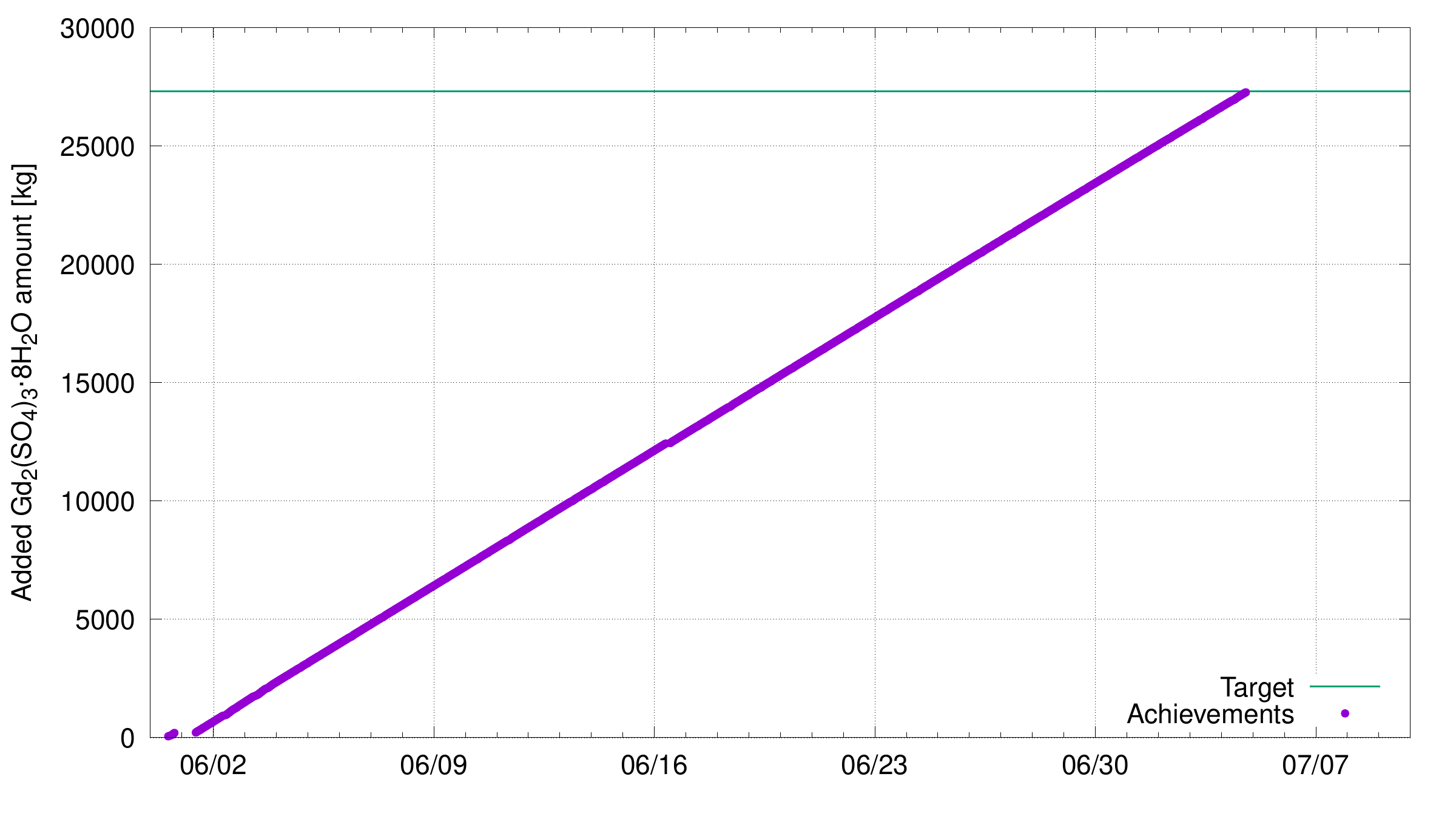}%
    \caption{A plot showing that a history of the total weight of Gd sulfate powder added into the SK water. The horizontal line colored in light blue indicates the goal of the second loading (27.3 tonnes). We achieved it on July 4th, 2022. \label{fig:gdload}}
\end{figure}

The second Gd loading took place from May 31 to July 4, 2022, i.e., it was a 36 day operation in total. 
The accumulated mass of the supplied Gd sulfate powder was 27,304~kg, which includes 4.5\% residual water. As the Gd fraction of \GdSOw is 0.421 based on the stoichiometric relationship, the corresponding mass of Gd is 10,998~kg.
The dissolution process took approximately 25 minutes to dissolve each 17~kg batch of powder, indicating a doubling of the dissolving rate compared to the initial loading, where 8.7~kg per batch was dissolved, as reported in the first loading paper~\cite{Gd1st}.
Every 8 hours, expert shift workers refilled the reserve tank with up to 400~kg of Gd sulfate powder. The batch numbers of the powder were recorded as part of the refilling work.
Thus, we can assess their radio-impurity contributions based on the screening results.
Figure~\ref{fig:gdload} shows a history of the total weight of Gd sulfate powder supplied to the feeder. It can be concluded that the loading work went smoothly during the entire period from the constant slope shown in the plot; a one-day overhaul for pump maintenance on June 17, 2022, can just barely be resolved.
The estimated total mass of Gd in the SK water following the second loading was 16,412~kg after taking loss of Gd during the work and the existing Gd from the first loading into consideration.
Therefore, the Gd concentration was $0.0332 \pm 0.0002$\% as a result of the second loading.
We conservatively assigned 0.5\% uncertainty on the masses of Gd and water. The uncertainly of water volume is mainly due to the expansion of the detector during water filling.

\subsection{Water transparency}
\begin{figure}
    \centering
    \includegraphics[width=0.98\textwidth]{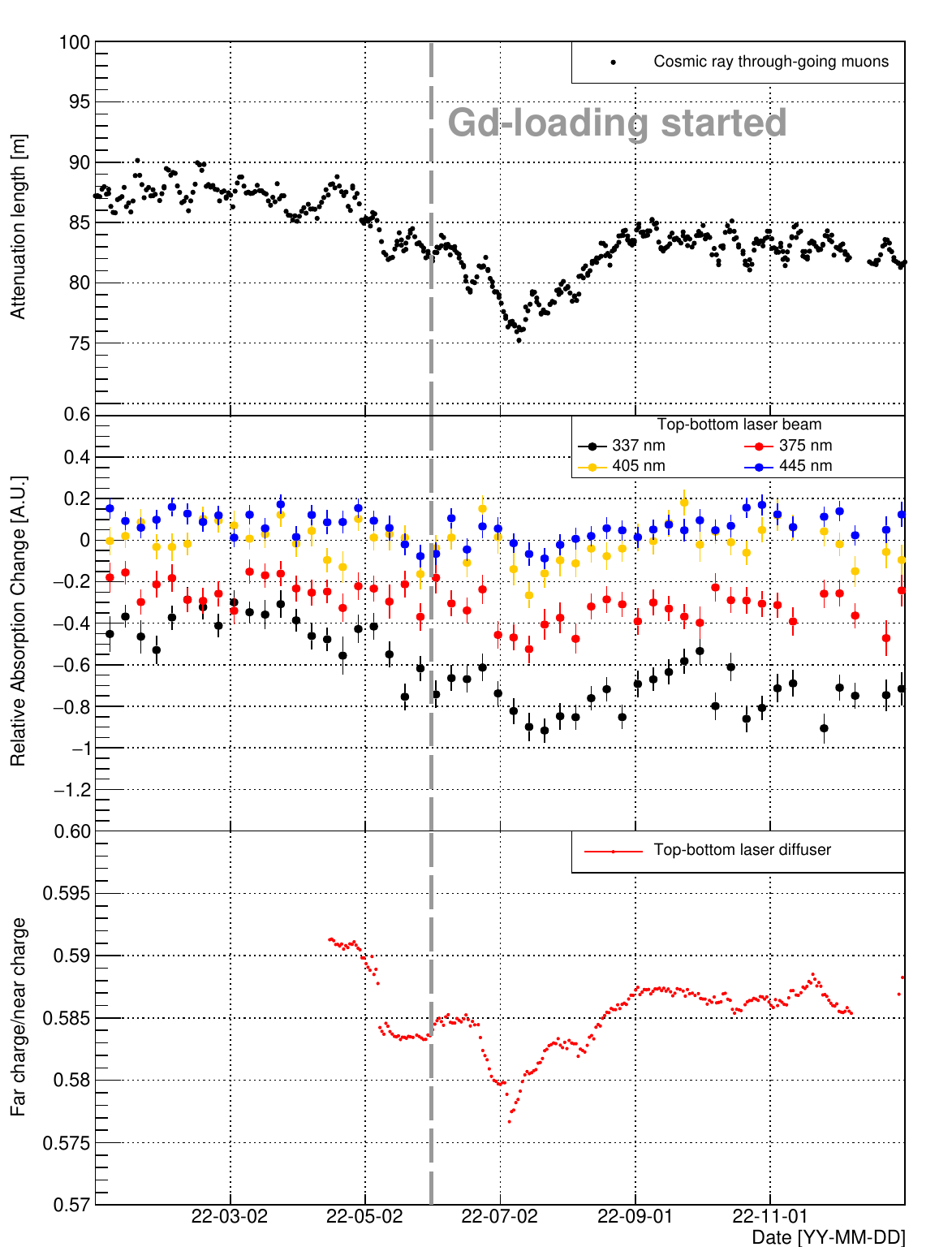}
\caption{
Time variation of the attenuation length for (top) through-going cosmic ray muons, (middle) top-bottom laser beam sources and (bottom) top-bottom laser diffuser, from January 1, 2022, to January 1, 2023. The larger decline in water quality observed around June-July 2022 corresponds to the Gd-loading period. Data from the top-bottom laser diffuser begins in April 2022 due to the installation of a new laser source, and the gap in November 2022 was caused by a loss of power to the laser. It is clear that the lower wavelength sources are more sensitive to changes in water parameters.\label{fig:watert}
}
\end{figure}

Time variation of the attenuation length of Cherenkov light measured in the SK tank using cosmic ray through-going muons around the second Gd loading period is shown in the top of Figure~\ref{fig:watert}. Details on this muon data analysis are described in~\cite{SKdet}, 
and the time variation of the attenuation length in the period before and after the first Gd loading in 2020 is described in~\cite{Gd1st}.
After the first Gd loading, the attenuation lengths recovered and became stable at $\sim90$~m with the water re-circulation system. In April 2022, the water re-circulation stopped several times for the maintenance of the water system, and the attenuation lengths dropped a little to $\sim85$~m.

Then, after May 6th, 2022, the water convection was intentionally started to raise the water temperature in preparation for the second Gd loading; convection made the attenuation length shorter. An additional  decrease in attenuation length was observed during the actual Gd loading phase starting from May 30th. 
When the loading finished, the attenuation length reached a minimum value of  $\sim$75~m, almost the same as what was seen following the first loading period.
Also, as we observed in the first loading, thanks to continuous operation of the water re-circulation system the attenuation length recovered within about two months after completion of the second Gd loading. 

In addition to the cosmic ray muon data, the laser beam data from the top to bottom of the tank (see~\cite{SKcalib}) and the diffuse laser data from top to bottom of the tank are also used to monitor the attenuation length in real time as shown in the middle and the bottom of Figure~\ref{fig:watert}. As real-time monitors where the time variation of a value is of interest, data from these sources are not converted to physics quantities. The PMT hit-time distributions after the prompt peak from the laser beam data are compared to a defined reference run with stable detector conditions. The slope of this distribution is sensitive to water quality, with improved transmission resulting in slope values greater than zero, and so is used as a real-time monitor of the change in water quality. The data from this source is shown in Figure~\ref{fig:watert}~(middle), with the reference period defined as before the first Gd loading period.
The diffuse laser light is produced by passing 368~nm laser light through a PTFE diffuser ball mounted at the top of the tank, resulting in a uniform cone of light with an opening angle of 40$^\circ$. A two-bin attenuation measurement is performed by taking the ratio of charge in a far to near region of the detector with respect to the diffuser position. Although a simple first analysis, the much greater statistics of this source provide a more precise measurement, which is capable of monitoring the time variation over shorter periods compared to the other two sources. This can be seen in Figure~\ref{fig:watert}~(bottom). As a result of these additional sources, it is found that the time variation of the attenuation length mainly comes from the absorption of shorter wavelengths ($\sim$340~nm) of light in water. In addition, especially at the initial stage of Gd loading, laser data and cosmic ray muon data at the bottom of the outer detector (OD) were used to monitor the effect on the detector in real-time.

There is a continuous gradual decrease in attenuation length from September 2022. The reason could be the reduction of water flow in the ID region of the tank. To further suppress the convection region and enlarge the low background region~\cite{Gd1st}, the water supply to the inner detector bottom region is limited to 28 m$^3$/h out of the total 120 m$^3$/h re-circulation flow (with the remainder going to the OD bottom region), and this causes water stagnation and worse water transparency.
Before the first Gd-loading, we modeled light attenuation and scattering based on EGADS data from a period when it was operated with a 0.1\% Gd concentration~\cite{EGADS}.  This EGADS-derived model was introduced into the SK detector simulator to generate the cosmic-ray muon events and perform the same muon data analysis as used here. The attenuation length with 0.1\% Gd in EGADS was measured to be $\sim67$~m. Next, assuming this water quality, physics events of interest were simulated in both the high-energy atmospheric neutrino analyses and the low-energy solar neutrino  analyses, and it was concluded that the impact on those physics analyses is acceptable. Therefore, 67~m is one guideline for the minimum required attenuation length. The current attenuation length in SK is significantly longer than this guideline, so there is no problem, but if eventually deemed necessary then transparency can be improved by increasing the water flow going to the ID region.

\section{Gd concentration measurements}
\subsection{Sampling water}
Similar to the first gadolinium loading, periodic samples were collected from calibration ports in the SK detector during the second loading. From May 31st to July 4th, 2022, samplings were conducted twice a week, approximately every 3 or 4 days. During the first loading, two ports in the ID and two ports in the OD were used. However, since no differences were seen in Gd concentration at any time within the ID and within the OD, one port in the ID and one in the OD were used for the second loading.

Following the first and second loadings, monthly samplings have been performed to assess the homogeneity and Gd concentration in the SK tank. Although the sampling system has undergone some maintenance whenever necessary, its structure has remained unchanged. It consists of a sampling probe comprised of a 25~cm stainless steel tube at the tip of a 50-meter sampling tube, and the sampling system, which, upstream from the SK tank, includes a flow meter, a pump,  and a device with dual output for temperature and conductivity measurements. After this, there is a port where samples can be taken. Most water circulating through this system is sent back to the SK water system. Only the water used to flush the sampling system is discarded, which is a total of about 15 liters per day of sampling (which includes sampling from the ID and OD ports). Samples of about 10~mL each are collected and later analyzed using an Atomic Absorption Spectrometer (AAS).

\begin{figure}[htbp]
    \centering
    \includegraphics[width=1.0\textwidth]{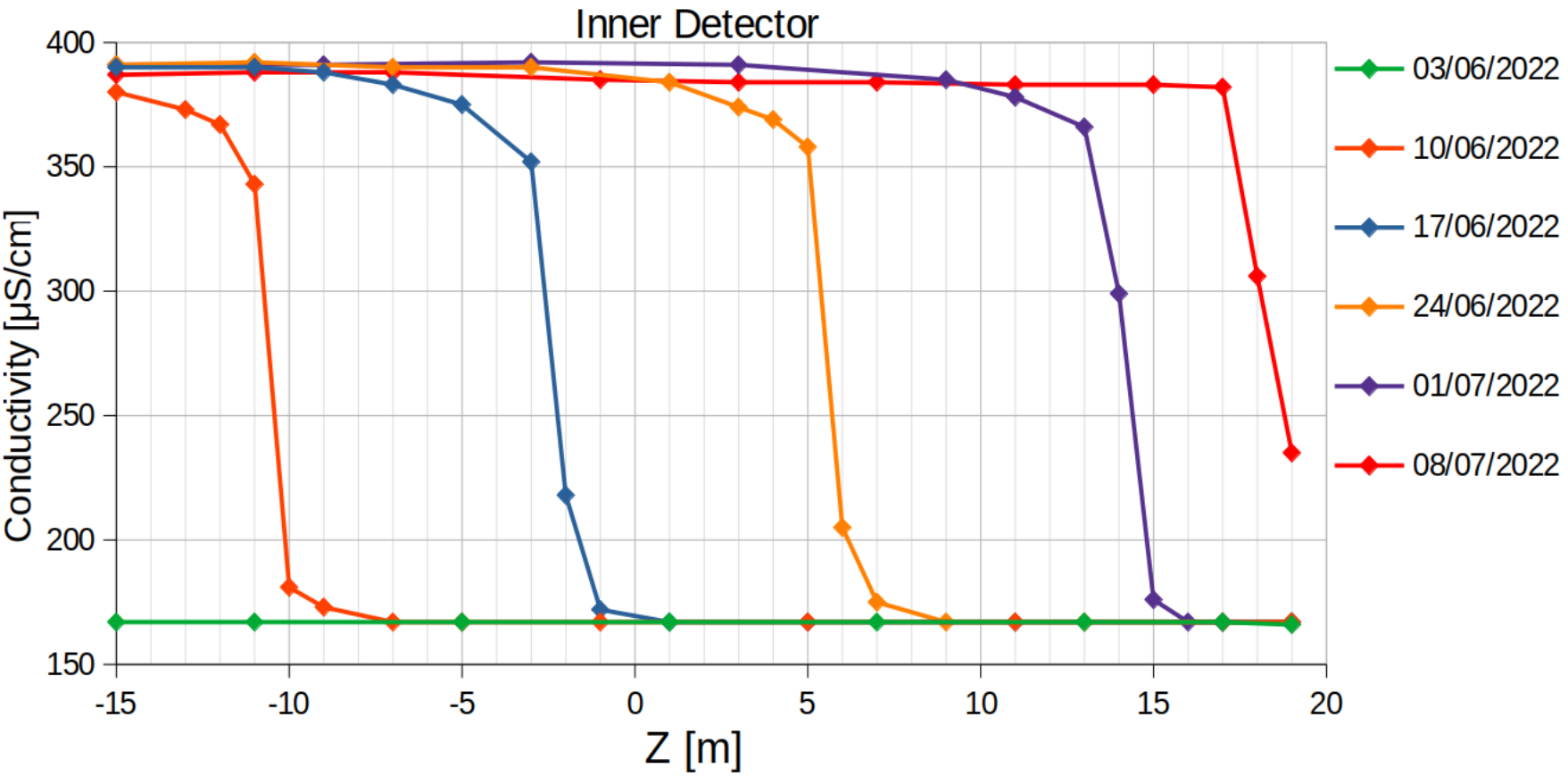}%
    \caption{Time evolution of conductivity in the ID region as a function of $Z$. $Z>+18$~m region is affected by the calibration pipe for the ID. \label{fig:IDConductivity}}
\end{figure}

Before the second loading, the Gd concentration was homogeneous throughout the detector at 113 $\pm$ 1 ppm. The conductivity was 167 $\mu$S/cm, which is a good indicator of the gadolinium sulfate concentration since the water purification systems are designed to remove all impurities except for gadolinium and sulfate ions. Once Gd-loaded water started being injected into the bottom of the SK tank, both conductivity and Gd concentration started increasing. Figure~\ref{fig:IDConductivity} shows the time evolution of conductivity in the ID region as a function of $Z$. On June 3rd, 2022, there was no change in conductivity in the ID yet, but a sharp rise was observed in the following days.  As in the first loading, a relatively narrow boundary about 2~m thick existed between the newly injected Gd at the bottom and the previous Gd concentration in the higher regions of the detector. This boundary advanced upwards at a rate slightly greater than 1~m/day. 
On July 8th, a few days after the Gd injection had been completed, the conductivity was almost homogeneous.
Note that $Z=+18.1$~m is at the inner surface of the ID while  $Z=+19$~m is inside of the pipe of the calibration port.

\begin{figure}[htbp]
    \centering
    \includegraphics[width=1.0\textwidth]{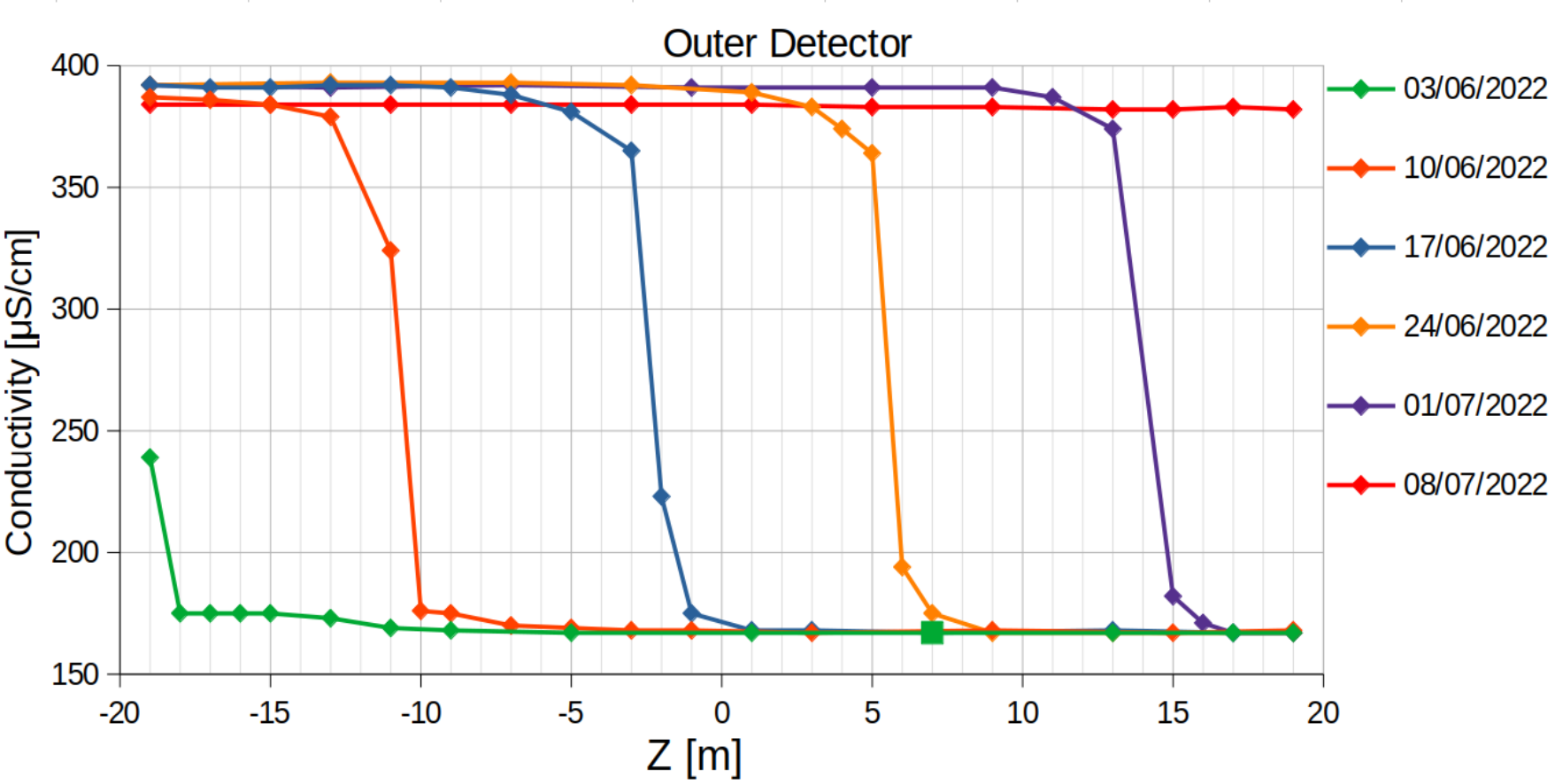}%
    \caption{Time evolution of conductivity in the OD region as a function of $Z$. \label{fig:ODConductivity}}
\end{figure}

Figure~\ref{fig:ODConductivity} shows the time evolution of the conductivity in the OD region as a function of $Z$. The boundary thickness and speed are similar to that of the ID, but, due to the larger water flow in the OD, the boundary was about 1~m higher than in the ID. On July 8th, 2022, the conductivity in the OD was homogeneous.

The collected samples were analyzed with an AAS. Standard samples of 20~ppm and 10~ppm were made with the same Gd sulfate octahydrate powder that was used for the loading. The uncertainty of a single measurement was about 2$\%$~\cite{martimagro2023gadolinium}. The measurements with the AAS show that the concentration of Gd sulfate octahydrate is homogeneous in the SK detector with an average value of 791.5~$\pm$~5.5~ppm. To translate this concentration to a concentration of Gd only, the non-Gd components have to be taken into account according to the stoichiometric relationship of gadolinium sulfate octahydrate. This yields a concentration of 333.5~$\pm$~2.5~ppm of Gd.

The absolute Gd concentration was also directly measured. For that, a Gd 1000 ppm commercial standard sample with a 2$\%$ nitric acid matrix was used. This standard sample was diluted to 8.42 ppm to perform this measurement with the 
AAS. We observed that the presence of nitric acid reduced the measured absorption by the AAS, thus artificially reducing the measured concentration. To account for this effect, we added nitric acid in the sample taken from the detector to a concentration such that after diluting it for the measurement, the nitric acid concentrations of the standard sample and the sample from the detector were the same. The AAS measurement yields a concentration of 332 $\pm$ 4 ppm of Gd.

\subsection{Neutron capture}
As natural gadolinium has a thermal neutron capture cross-section $10^5$ times larger than that of hydrogen ~\cite{ENDF7,IUPAC2016},
there's a direct relationship between the neutron capture time constant and the concentration of gadolinium in the water.
Therefore, Gd concentration was also evaluated by observing neutron captures on Gd. We utilized two neutron sources: spallation neutrons and neutrons from an Am/Be source. Spallation neutrons are neutrons produced by cosmic ray-induced spallation reaction of ${}^{16}$O in the detector medium; their advantage is the capability to monitor the entire detector volume continuously.
The Am/Be source is more suitable for precisely determining the Gd concentration thanks to the well-defined neutron emission process.
We describe the results of Gd concentration measurements using these two methods.

\subsubsection{Spallation neutron}
\label{sec:spa}
Selection of spallation neutrons is done using a method similar to the one described in Ref.~\cite{Super-Kamiokande:2022cvw}. Cosmic muons passing through the SK detector were selected by requiring both SHE (Super High Energy) and OD triggers to be issued. An SHE trigger is issued with more than 60 ID PMT hits within a 200~ns time window, and an OD trigger is issued with more than 22 OD PMT hits also within a 200~ns window. An SHE trigger records all the PMT hits within 35 ~$\mu$s after the trigger timing, followed by an AFT (after) trigger which records all the hits within an additional 500~$\mu$s time window. Then neutron candidates are selected by requiring more than 20 ID PMT hits within a 200~ns time window from the hits recorded by the SHE and the AFT triggers. Since SK observes $\sim$6 Chernekov photo-electrons per MeV, this corresponds to roughly a 2.5~MeV electron equivalent once dark noise hits are taken into account. 
Further selection on the analysis volume, number of hits within a 50~ns time window, reconstruction quality, and the distance between muon track and neutron candidates are applied to reduce backgrounds. 
The selected neutron candidates are predominantly from neutron
captures on Gd, with the contamination of captures on hydrogen
estimated to be no greater than 0.4\%.

\begin{figure}[htbp]
    \centering
    \includegraphics[width=0.9\textwidth]{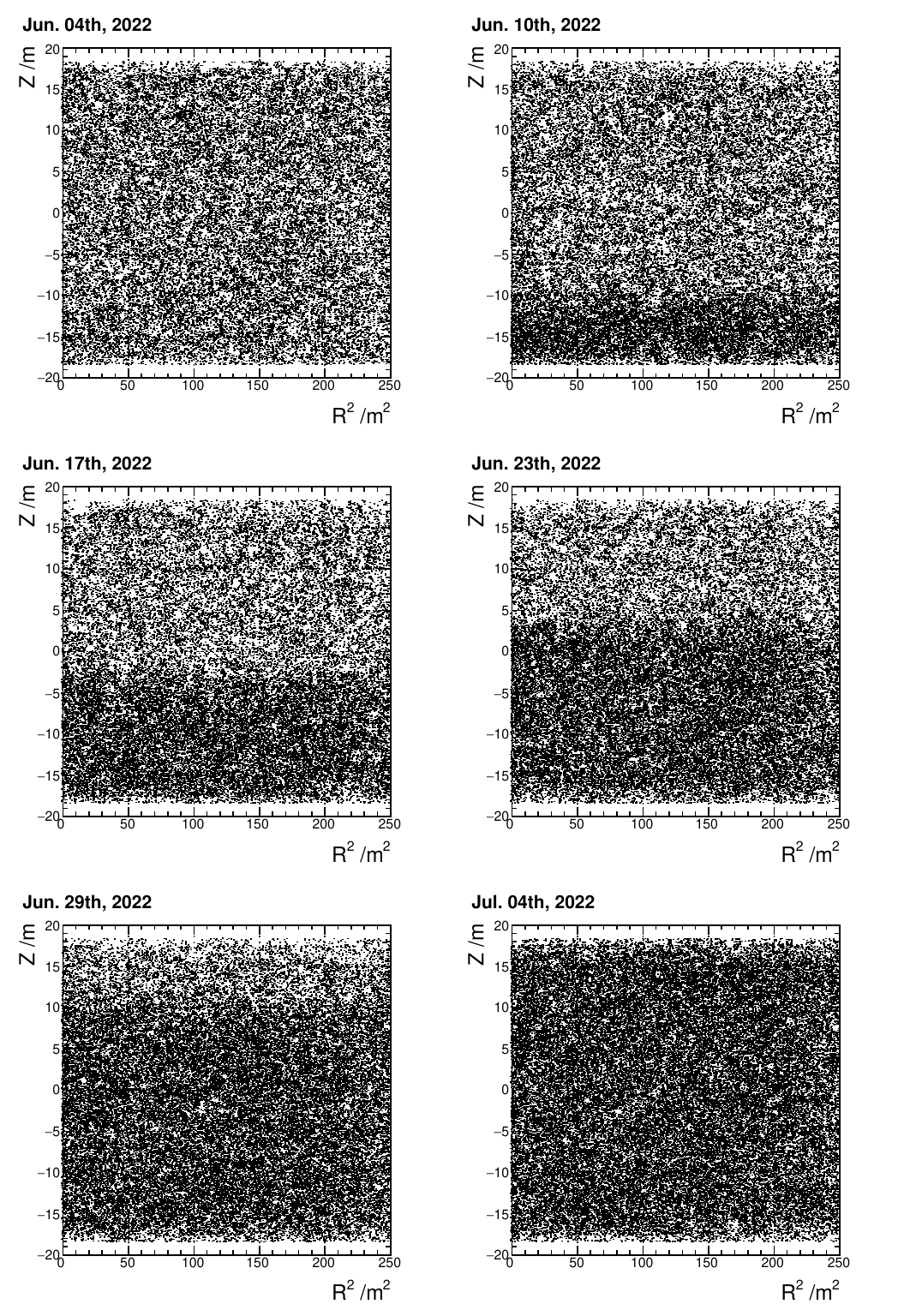}%
    \caption{Reconstructed vertex distributions in the SK tank of spallation neutron candidates during the Gd loading from May 31st to July 4th, 2022. The horizontal axis is reconstructed $R^2 = X^2 + Y^2$  and the vertical axis is the reconstructed Z position. Each panel shows the distributions from different periods. 
    \label{fig:vertex_spaln}}
\end{figure}

Figure~\ref{fig:vertex_spaln} shows the spatial distributions of spallation neutron candidates for each week after the start of the second Gd loading on May 31st, 2022. It clearly shows that the region with a higher density of neutron candidates gradually expands from the bottom of the tank to the top at a rate consistent with the water recirculation rate.

Measurements of the neutron capture time constant, which characterizes the time between prompt muon and delayed neutron capture events, are used to monitor the change in the Gd concentration.
The capture time constant can be described by the number density of H and Gd ($n_{\mathrm{H, Gd}}$) and their capture cross sections ($\sigma_{\mathrm{H, Gd}}$) as follows:
\begin{equation}
    \tau = \frac{1}{\sum_{i=\mathrm{H, Gd}} n_{i} g_{\mathrm{w}}^{i} \sigma_{i}^{therm} v_{therm}},
\end{equation}
where $g_{\mathrm{w}}$ is the Wescott g-factor (constant), $v_{therm}$ is the speed of a thermal neutron (2200 m/s), and $\sigma^{therm}$ is an averaged cross section over a Maxwell distribution with $v_{therm}$.

\begin{figure}[htbp]
    \centering
\includegraphics[width=0.9\textwidth] 
    {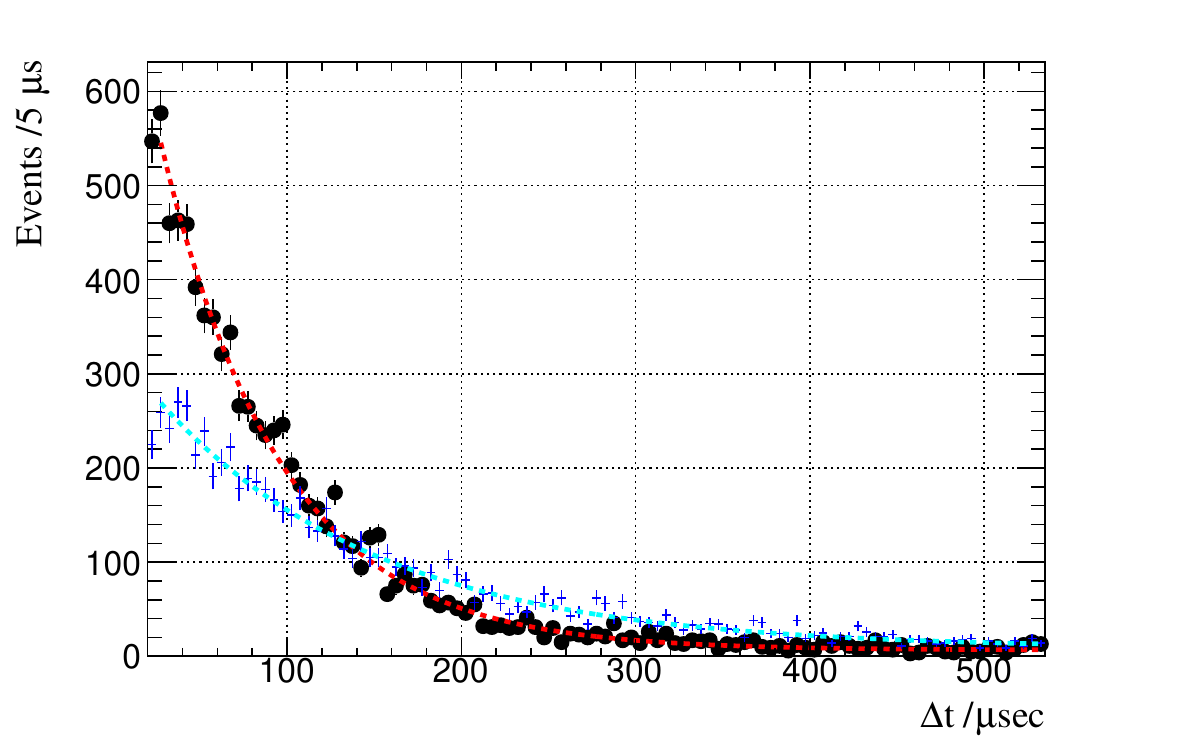}%
    \caption{A typical distribution of the timing difference between muons and their spallation neutron candidates. The data was obtained on June 19th, 2022, when the border between high and low Gd concentration regions was around ${\rm Z} = 0$~m. The filled markers show the data points with a selection of $-16.10 < {\rm Z} < -12.88$~m corresponding to $\sim0.033\%$ Gd concentration. The red line shows a fitted function of a single exponential plus a constant background. The blue crosses and cyan dashed line show the corresponding data points and a fitted curve before the second Gd-loading ($\sim0.011\%$ Gd concentration) with a selection of $+12.88 < {\rm Z} < +16.10$~m. \label{fig:dt_spaln}}
\end{figure}

\begin{figure}[htbp]
    \centering
    \includegraphics[width=1.0\textwidth]{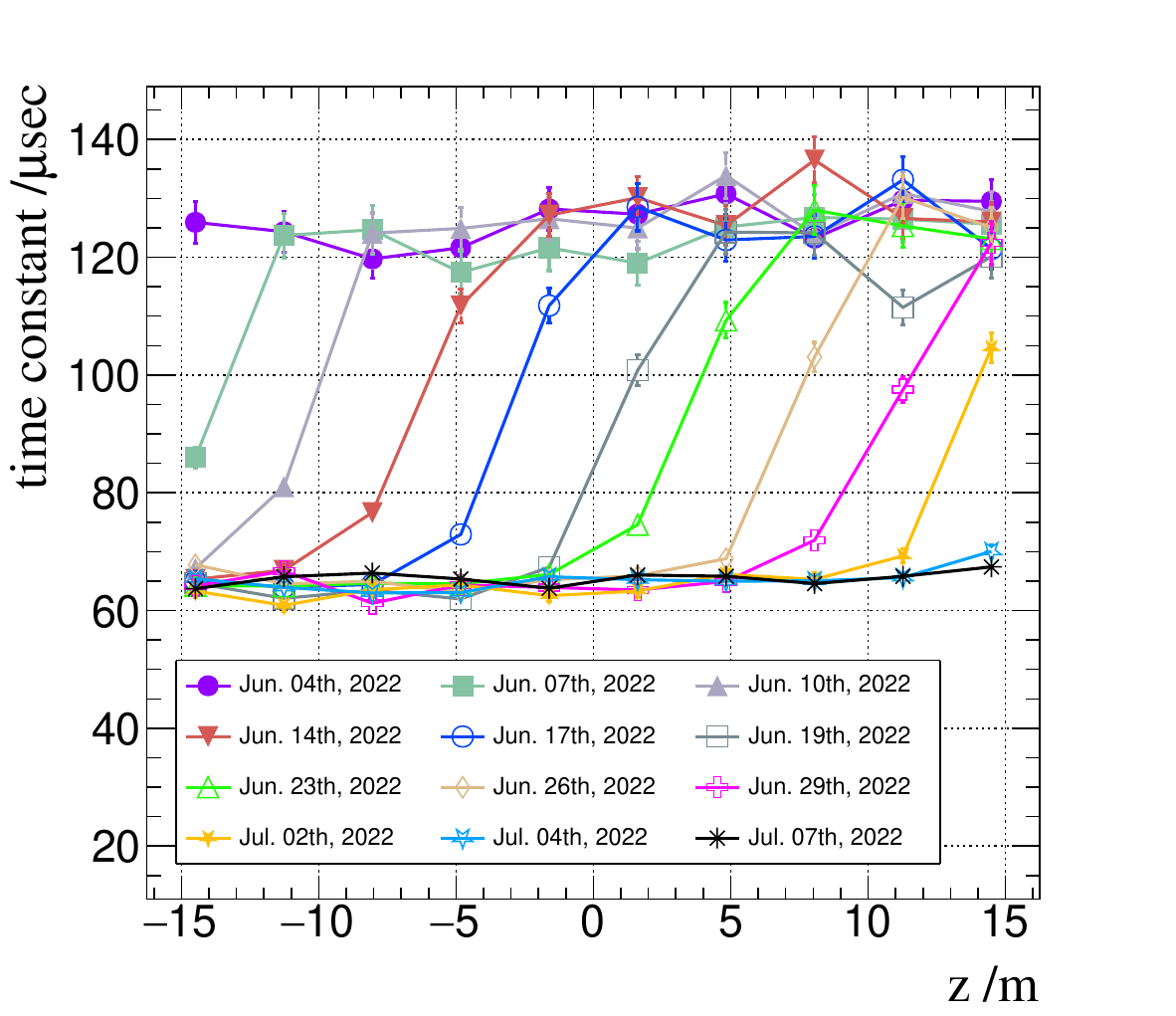}%
    \caption{Capture time of spallation neutrons during the second Gd loading, evaluated in $\sim$3-day periods at 10 positions dividing the ${\rm Z}$ axis into 3.22~m-thick slices. The mean of $\chi^2 / d.o.f.$ is 1.02 and its standard deviation is 0.16 for these fits. Thus, the fits give sufficiently good description on the data. \label{fig:tcap_spaln}}
\end{figure}

Figure~\ref{fig:dt_spaln} shows a typical distribution of the time difference between muon and neutron capture candidates. 
The time constant was extracted by fitting this distribution with a function with a single exponential decay plus a constant background.
The capture time is expected to change from $\sim$120~$\mu$s at 0.01\% Gd to $\sim$60~$\mu$s at 0.03\% Gd.
Figure~\ref{fig:tcap_spaln} shows extracted capture time
constants evaluated for each $\sim$3-day period, with each such sub-sample separated by the reconstructed vertical (${\rm Z}$) positions of the neutron candidates. It clearly shows a transition from 0.01\% to 0.03\% Gd concentration from the bottom to the top of the detector, consistent with the expectation. 

\subsubsection{Am/Be neutron}
Like the first Gd loading, an 
americium-beryllium (Am/Be) neutron source $({}^{241}\!\mathrm{Am}\rightarrow {}^{237}\mathrm{Np}+\alpha, {}^{9}\mathrm{Be}+\alpha\rightarrow {}^{12}\mathrm{C}^{*}+\mathrm{n}, {}^{12}\mathrm{C}^{*}\rightarrow{}^{12}\mathrm{C}+\gamma\mathrm{(4.4~MeV)})$ was also deployed this time as another way of measuring the neutron capture time constant~\cite{Gd1st}.
The Am/Be source is encapsulated with 5$\times$5$\times$5~cm bismuth germanate (BGO) scintillator crystal cubes, whose light results in  $\sim$1000 scintillation photoelectrons for the detection of the full-energy peak of 4.4~MeV gamma-rays as the prompt signal.
Subsequent neutron capture on gadolinium is identified by its gamma-ray emission from the excited capture nucleus.
Such gamma rays are detectable through Compton scattering off electrons in the SK water, producing Cherenkov light. 
The total energy of gamma rays from neutron captures on gadolinium is typically $\sim$8~MeV, which primarily comes from two isotopes, $^{155}$Gd and $^{157}$Gd :
	$\mbox{n}+{}^{155}\mbox{Gd}\rightarrow {}^{156}\mbox{Gd}+\gamma\mbox{'s}\ [\mbox{8.5~MeV\ in\ total}],
	\mbox{n}+{}^{157}\mbox{Gd}\rightarrow {}^{158}\mbox{Gd}+\gamma\mbox{'s}\ [\mbox{7.9~MeV\ in\ total}]$.

The data were taken through calibration ports near the center in the X-Y plane, $(x,y)=(-0.4,-0.7)$ and $ (-3.9,-0.7)$~[m].
Three positions along the Z-coordinate were selected for periodic monitoring: $z=0, +12$ and $-12$~m.

As explained in Section~\ref{sec:spa}, 
for neutron data taking with the Am/Be source the SHE and following AFT triggers are applied, though
in this case more than 100 ID PMT hits (not 60) within a 200~ns time window are required to issue an SHE trigger.
Gd(n,$\gamma$)Gd event candidates are extracted from the recorded PMT data by looking for more than 30 ID PMT hits in a 200~ns time window and applying event vertex reconstruction.

\begin{figure}[htb]
	\centering\includegraphics[width=0.8\linewidth]{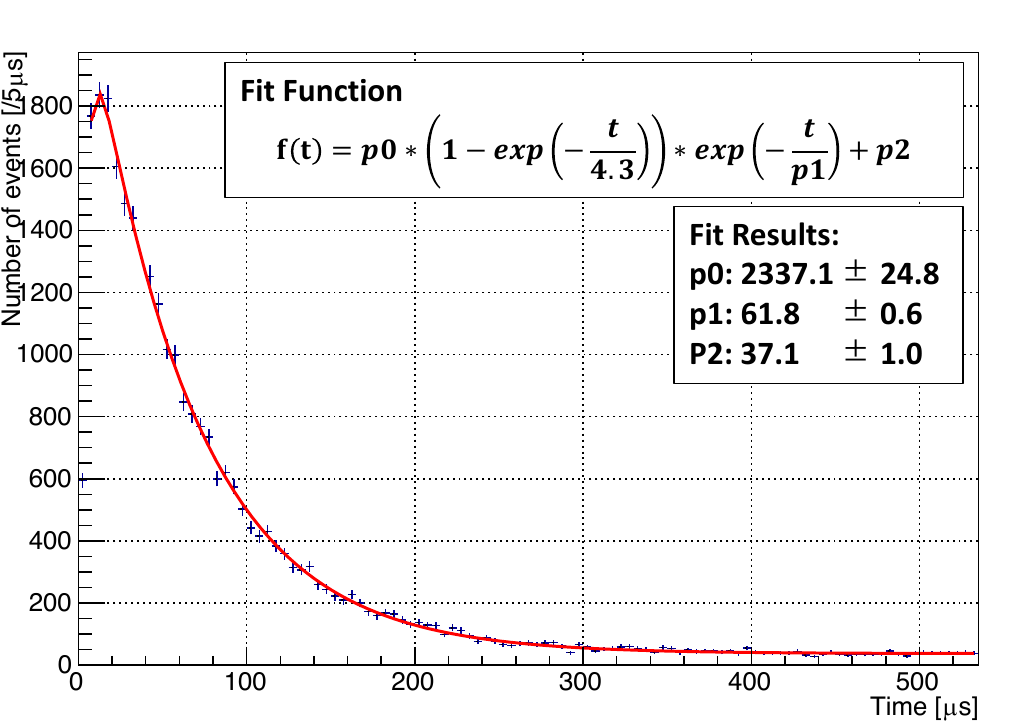}
	\caption{An example of the time distribution of neutron capture event candidates (black data points) and its fit function (red line), measured with the Am/Be source at the Z=0~m position on September 28, 2022.  Time zero is defined by the detection of the prompt 4.4~MeV gamma-ray BGO scintillation event. 
	The neutron capture time constant is represented by $p1$, while the thermalization time constant of 4.3~$\mu$s is derived from a summed analysis of these measurements. 
	}
	\label{fig:TimeSpectrum_28thSep2022}
\end{figure}

A typical time distribution of neutron capture event candidates is shown in Figure~\ref{fig:TimeSpectrum_28thSep2022}.
 The analysis applied event selection criteria to the SK event reconstruction parameters. Specifically, the reconstruction timing goodness was required to be greater than 0.4, the hit pattern goodness had to be smaller than 0.4, and the event vertex had to be located within 4~m from the Am/Be source position in the SK tank~\cite{Abe:2016nxk}.
For the initial SHE trigger events, it was required that they occurred at least 1ms after the previous SHE trigger to mitigate interference from the neutron emitted in the prior event. Additionally, these initial SHE trigger events had to encompass 850 to 1250 active PMT hits within a time window of 1.3$\mu$s. This selection ensured the isolation of the 4.4~MeV gamma-ray emission originating from the Am/Be source.

\begin{figure}[htb]
	\centering\includegraphics[width=1.0\linewidth]{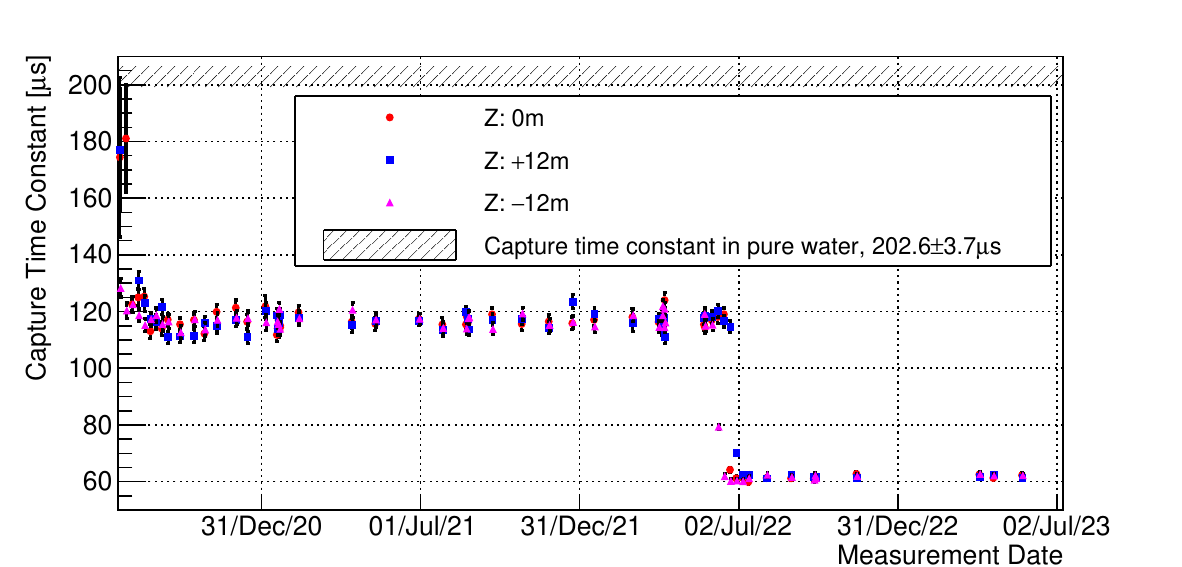}
 	\centering\includegraphics[width=1.0\linewidth]{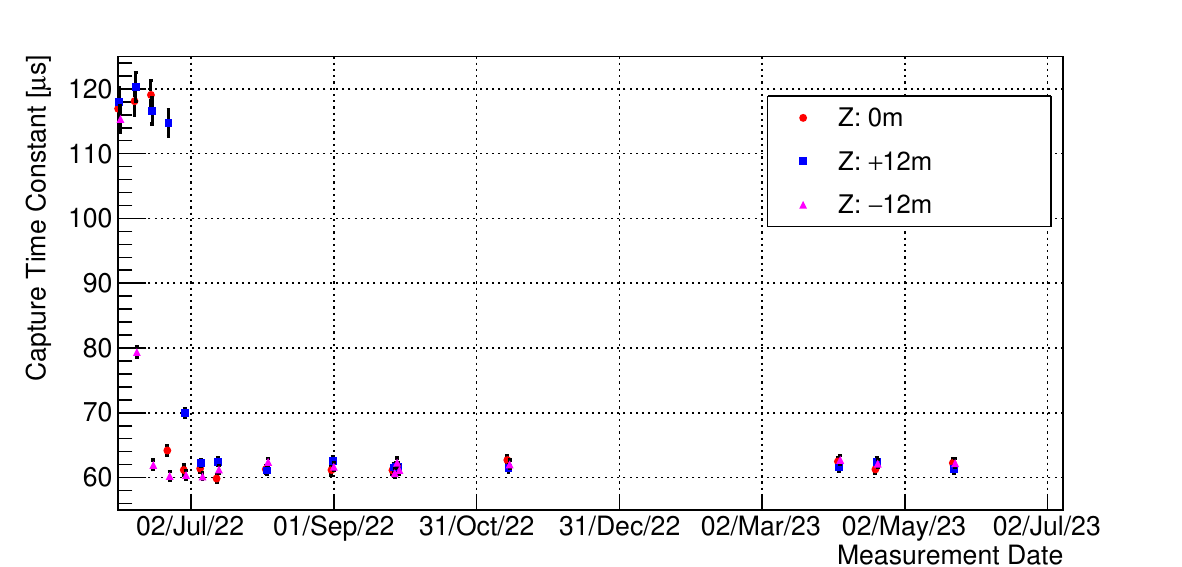}
	\caption{The history of the neutron capture time constant in SK since July 2020 (top) and after this second Gd-loading period (bottom), obtained from the analysis of Am/Be source data. Data were taken using a calibration port at the detector's center (X=$-$0.3~m, Y=$-$0.7~m) and near the center in the X-Y plane (X=$-$3.9~m, Y=$-$0.7~m).
	Three positions along the Z-coordinate were selected for periodic measurement: Z=0~m (red circles), Z=$+$12~m (blue squares) and Z=$-$12~m (magenta triangles).
	The shaded area indicates the neutron capture time constant in pure water
	\cite{zhang:2016prd}.
	}
	\label{fig:History_Of_GdCaptureTime}
\end{figure}

In fitting the event candidate time distribution, neutron thermalization and capture time constants were considered as well as the presence of background events, which exhibited a constant distribution in time.
As shown in Figure~\ref{fig:History_Of_GdCaptureTime},
the neutron capture time constant demonstrated reasonable stability over both depth within SK and time from August 2022 to May 2023.
The mean neutron capture lifetime of 61.8~$\pm$~0.1~$\mu$s is given by Gaussian fitting to these 27 data points.
The standard deviation of lifetime was 0.5$\pm$~0.1~$\mu$s, which is well explained by the uncertainty of each Am/Be data-taking run.

\begin{figure}[htb]
	\centering\includegraphics[width=0.8\linewidth]{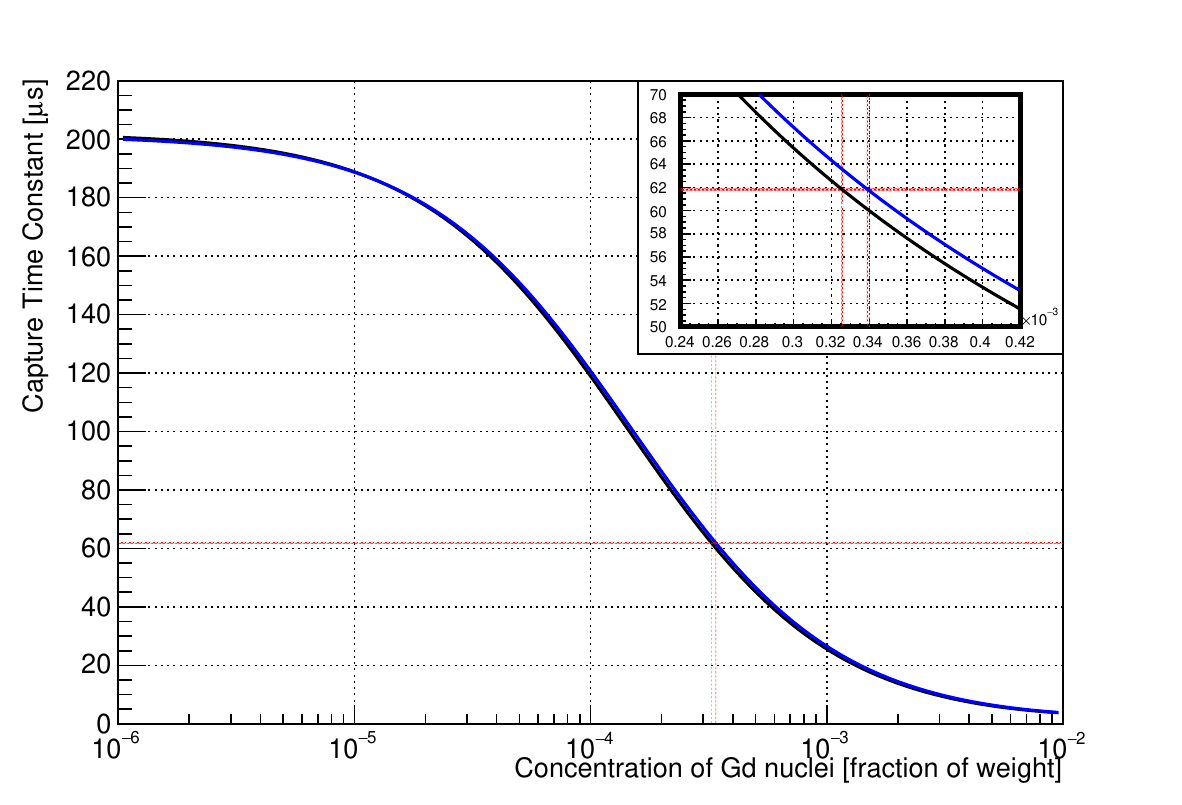}
	\caption{Neutron capture time constant as a function of the gadolinium concentration. The black line correspond to an approximation function, given by fitting the Geant4.10.5p01 and G4NDL 4.5 Monte Carlo simulation.
    The blue line is another approximate polynomial function, for the Geant4.9.6p04 and G4NDL4.2 simulation.
    The geometrical configurations of the simulations are identical to  what was in our previous paper~\cite{Gd1st}.
The horizontal and vertical red band represents the mean neutron capture time constant and derived concentration.  
	}
	\label{fig:CaptureT_Gd_MC}
\end{figure}

To convert from capture lifetime to Gd concentration, Geant4-based Monte Carlo simulation results, as illustrated in Figure~\ref{fig:CaptureT_Gd_MC}, were applied in the analysis.
There are two evaluated Gd concentrations, 339.5~$\pm$~0.8~ppm (when using Geant4.9.6p04) and 325.9~$\pm$~0.7~ppm (when using Geant4.10.5p01).
If we consider the difference between these two evaluation values as the systematic error of the models, we can say 332.7~$\pm$~6.8(sys.)~$\pm$~1.1(stat.)~ppm.  This concentration is consistent with both the concentration derived from the gadolinium mass as well as the concentration measured by AAS. 

\color{black}

\color{black}

\color{black}

\section{Conclusion}
The introduction of 12.9 tons of Gd$_2$(SO$_4$)$_3\cdot$8H$_2$O in 2020, plus an additional 26.1 tons of Gd$_2$(SO$_4$)$_3\cdot$8H$_2$O in the summer of 2022, brings the dissolved Gd concentration in Super-Kamiokande to 0.033\%.
For this second loading in particular -- to allow the continuation of SK's solar neutrino observations at low energy -- we developed gadolinium sulfate with even fewer radioactive impurities than that used in the first loading. All the production batches were then screened to confirm their purity.
As in the first loading, using the density difference of the solutions we successfully collected 0.011\% Gd solution from the top of the tank and introduced 0.033\% Gd solution from the bottom of the tank. During this process, the Gd concentration was continually monitored using the capture time constant of spallation neutrons produced by cosmic ray muons. The water transparency variation was also tracked, and from the laser beam data we confirmed that the shorter wavelength sources ($<400$ nm) are more sensitive to changes in water transparency.
From the measurement of Gd concentration by AAS and the time constant of neutron capture using Am/Be neutron sources, it was confirmed that the Gd concentration became uniform in the tank just after the second Gd loading, and the neutron capture efficiency by Gd increased by 1.5 times as expected.
The second gadolinium loading to SK-Gd is expected to allow it to make the world's first observation of the DSNB flux within a few years, and more accurate supernova burst direction determination via inverse beta decay identification in the case of a galactic supernova explosion.

\section*{Acknowledgment}
We gratefully acknowledge the cooperation of the Kamioka Mining and Smelting Company.
The Super‐Kamiokande experiment has been built and operated from funding by the 
Japanese Ministry of Education, Culture, Sports, Science and Technology, the U.S.
Department of Energy, and the U.S. National Science Foundation. Some of us have been 
supported by funds from the National Research Foundation of Korea NRF‐2009‐0083526
(KNRC) funded by the Ministry of Science, ICT, and Future Planning and the Ministry of
Education (2018R1D1A3B07050696, 2018R1D1A1B07049158), 
the Japan Society for the Promotion of Science (JSPS KAKENHI Grant Numbers JP19H05807,JP26000003), the National
Natural Science Foundation of China under Grants No.11620101004, the Spanish Ministry of Science, 
Universities and Innovation (grant PGC2018-099388-B-I00), the Natural Sciences and 
Engineering Research Council (NSERC) of Canada, the Scinet and Westgrid consortia of
Compute Canada, the Ministry of Science and Higher Education (2023/WK/04) and the National Science Centre (UMO-2018/30/E/ST2/00441 and UMO-2022/46/E/ST2/00336), Poland,
the Science and Technology Facilities Council (STFC) and GridPPP, UK, the European Union's 
Horizon 2020 Research and Innovation Programme under the Marie Sklodowska-Curie grant
agreement no.754496, H2020-MSCA-RISE-2018 JENNIFER2 grant agreement no.822070, and 
H2020-MSCA-RISE-2019 SK2HK grant agreement no. 872549.

\appendix
\section{Screening results}
\subsection{ICP-MS}
\label{apdx:ICP-MSres}


The results from the ICP-MS assays of all batch samples of \GdSOw are shown in Table~\ref{tab:ICPresult}. 
In this table, in order to make it easier to compare the quality of gadolinium sulfate used in the initial loading,
ppt for U and Th and ppb for Ce in the same unit as Table 4 in \cite{1stGdProdScrn} are used. The conversion factors are given as 1~ppt (U) $=1.24\times 10^{-2}$~mBq/kg and 1~ppt (Th) $=4.06\times 10^{-3}$~mBq/kg.

\begin{table}[!ht]
\vspace{-1.6cm}
    \centering
    \begin{tabular}{r|ccc}
    \hline
        Batch ID & U [ppt] & Th [ppt] & Ce [ppb]  \\ 
                  & $<$ 400 & $<$ 13  & $<$ 50 \\ \hline
210301 & 3.07±0.12 & 12.43±1.69 & 1.28±0.12 \\ \hline
210302 & 1.87±0.26 & 1.88±0.1 & 0.13±0.06 \\ \hline
210303 & 2.94±0.08 & 0.77±0.18 & 0.72±0.08 \\ \hline
210601 & 2.68±0.13 & 2.09±0.34 & 1.11±0.15 \\ \hline
210711 & 0.0±0.15 & 1.33±0.09 & 2.05±0.14 \\ \hline
210712 & 1.35±0.13 & 1.91±0.13 & 9.31±0.68 \\ \hline
210713 & 1.37±0.2 & 3.96±0.19 & 1.14±0.17 \\ \hline
210811 & 1.96±0.16 & 1.36±0.26 & 0.65±0.09 \\ \hline
210821 & 1.45±0.04 & 2.27±0.15 & 0.84±0.28 \\ \hline
210822 & 1.09±0.21 & 1.7±0.18 & 0.7±0.16 \\ \hline
210823 & 2.75±0.19 & 6.71±0.57 & 0.83±0.15 \\ \hline
210922 & 1.1±0.03 & 11.83±1.1 & 1.14±0.24 \\ \hline
211006 & 2.82±0.46 & 2.57±0.17 & 0.40±0.17 \\ \hline
211106 & 1.29±0.13 & 0.7±0.14 & 1.00 ±0.1 \\ \hline
211201 & 1.39±0.13 & 1.3±0.1 & 1.27±0.2 \\ \hline
211202 & 1.13±0.19 & 1.05±0.15 & 0.44±0.24 \\ \hline
211204 & 0.83±0.15 & 1.95±0.11 & 1.51±0.07 \\ \hline
211205 & 0.94±0.19 & 8.77±0.69 & 0.39±0.16 \\ \hline
220102 & 1.87±0.08 & 2.27±0.16 & 0.53±0.17 \\ \hline
220103 & 1.23±0.11 & 2.59±0.29 & 0.39±0.18 \\ \hline
220104 & 1.01±0.13 & 1.82±0.19 & 0.42±0.11 \\ \hline
220201 & 1.08±0.16 & 2.02±0.19 & 0.35±0.09 \\ \hline
220241 & 1.4±0.1 & 1.24±0.18 & 0.49±0.19 \\ \hline
220242 & 1.21±0.23 & 1.4±0.11 & 0.75±0.13 \\ \hline
220251 & 1.51±0.1 & 0.85±0.16 & 0.94±0.15 \\ \hline
220351 & 1.5±0.18 & 1.29±0.13 & 0.59±0.24 \\ \hline
220352 & 1.3±0.07 & 0.91±0.05 & 0.59±0.06 \\ \hline
220353 & 1.59±0.28 & 0.97±0.1 & 0.29±0.02 \\ \hline
220361 & 1.36±0.08 & 0.8±0.12 & 0.29±0.09 \\ \hline
220371 & 3.11±0.15 & 1.17±0.06 & 0.7±0.09 \\ \hline
220471 & 1.44±0.2 & 0.34±0.25 & 0.14±0.11 \\ \hline
220481 & 2.75±0.12 & 1.39±0.14 & 0.49±0.21 \\ \hline
220482 & 3.21±0.25 & 1.48±0.21 & 0.27±0.14 \\ \hline
220581 & 1.67±0.35 & 0.86±0.07 & 0.9±0.06 \\ \hline
220582 & 1.87±0.11 & 1.11±0.09 & 0.79±0.16 \\ \hline
220691 & 1.29±0.11 & 0.35±0.17 & 0.73±0.12 \\ \hline
220603 & 2.32±0.24 & 0.42±0.14 & 0.89±0.47 \\ \hline
    \end{tabular}
    \caption{\label{tab:ICPresult}
    Results from ICP-MS assays of all samples. The SK requirement on the concentration of each element is indicated at the top of each column.}
\end{table}

\subsection{HPGe}
\label{app:Ge}
Table~\ref{tab:GeResult2021} and Table~\ref{tab:GeResult2022} show  results from the HPGe measurements for all of the \GdSOw samples.

\begin{landscape}
\begin{table}[!ht]
    \centering
    \scalebox{0.78}[0.78]{
    \begin{tabular}{r|cc|ccccccccccc}
    \hline

Batch ID & Lab     & detector  & $^{238}$U & $^{226}$Ra & $^{228}$Ra & $^{228}$Th & $^{235}$U & $^{223}$Ra & $^{40}$K & $^{138}$La & $^{176}$Lu & $^{134}$Cs & $^{137}$Cs \\ \hline
210301   & Kamioka & LabC01    & $<$6.4    & $<$0.43    & $<$0.22    & $<$0.17    & $<$2      & $<$1.2     & $<$1.8   & $<$0.045 & 0.37$\pm$0.089& $<$0.065& $<$0.12 \\
210302   & LSC     & geAsterix & $<$10& $<$0.19& $<$0.37& $<$0.38& $<$0.47& $<$1.7& $<$1.8& $<$0.21 & 0.16$\pm$0.05& $<$0.09& $<$0.09 \\
         & Kamioka & LabC01    & $<$8.6& $<$0.47& $<$0.66& $<$0.28& $<$3.2& $<$1.2& $<$2.1& $<$0.08& $<$0.37& $<$0.087& $<$0.21 \\
210303   & LSC     & geOreoel  & $<$8& $<$0.34& $<$0.66& $<$0.36& $<$0.47& $<$1.9 & 2.0$\pm$0.6& $<$0.3 & 0.54$\pm$0.07& $<$0.13& $<$0.13 \\
         & Kamioka & LabC01    & $<$8.8& $<$0.45& $<$0.74& $<$0.24& $<$4& $<$1.3& $<$1.7& $<$0.062 & 0.78$\pm$0.11& $<$0.094& $<$0.18 \\
210601   & BUGS    & Belmont   & $<$13.66& $<$0.31& $<$0.39& $<$0.31& $<$0.24& $<$1.08& $<$2.14& $<$0.11 & 0.41$\pm$0.13 & -& $<$0.16 \\
         & Kamioka & LabC01    & $<$6.3& $<$0.25& $<$0.23& $<$0.17& $<$4.3& $<$0.75& $<$1.7& $<$0.048 & 0.62$\pm$0.095& $<$0.07& $<$0.13 \\
         & LSC     & geAsterix & $<$9& $<$0.16& $<$0.29& $<$0.29& $<$0.39& $<$1.4& $<$1.8& $<$0.10 & 0.46$\pm$0.05& $<$0.07& $<$0.10 \\
210711   & BUGS    & Merrybent & $<$4.86& $<$0.22 & 0.27$\pm$0.13& $<$0.31& $<$0.18& $<$0.99& $<$1.59& $<$0.14 & 0.41$\pm$0.06 & -& $<$0.04 \\
         & LSC     & geAsterix & $<$13& $<$0.26& $<$0.47& $<$0.48& $<$0.58& $<$2.3& $<$2.7& $<$0.22 & 0.20$\pm$0.06& $<$0.10& $<$0.11 \\
         & Kamioka & LabC01    & $<$4.9& $<$0.17& $<$0.28& $<$0.22& $<$2.2& $<$0.76& $<$1& $<$0.032 & 0.23$\pm$0.07& $<$0.056& $<$0.1 \\
210712   & LSC     & geAsterix & $<$15& $<$0.23& $<$0.47& $<$0.53& $<$0.67& $<$2.3& $<$1.8& $<$0.24 & 0.11$\pm$0.05& $<$0.09& $<$0.12 \\
         & BUGS    & Belmont   & $<$3.75& $<$0.33& $<$0.26& $<$0.18& $<$0.19& $<$0.83& $<$1.25& $<$0.10& $<$0.15 & -& $<$0.04 \\
         & Kamioka & LabC01    & $<$6.5& $<$0.17& $<$0.51& $<$0.18& $<$1.9& $<$0.7& $<$1& $<$0.039& $<$0.11& $<$0.06& $<$0.13 \\
210713   & BUGS    & Belmont   & $<$5.55& $<$0.26& $<$0.49& $<$0.34& $<$0.22& $<$1.08& $<$1.75& $<$0.12& $<$0.14 & -& $<$0.06 \\
         & Kamioka & LabC01    & $<$5.1 & 0.22$\pm$0.078& $<$0.26 & 0.17$\pm$0.083& $<$1.6& $<$0.61& $<$0.78& $<$0.033 & 0.16$\pm$0.06& $<$0.043& $<$0.11 \\
210811   & LSC     & geOreoel  & $<$10& $<$0.29& $<$0.50& $<$0.28& $<$0.39& $<$1.3& $<$1.4& $<$0.2 & 0.18$\pm$0.06& $<$0.12& $<$0.12 \\
         & BUGS    & Merrybent & $<$4.91 & 0.40$\pm$0.12& $<$0.24& $<$0.35& $<$0.34& $<$0.82& $<$1.98& $<$0.16 & 0.32$\pm$0.07 & -& $<$0.06 \\
         & Kamioka & LabC02    & $<$17& $<$0.3& $<$0.36& $<$0.2& $<$2.3& $<$0.97& $<$1.9& $<$0.094 & 0.21$\pm$0.09& $<$0.078& $<$0.12 \\
210821   & BUGS    & Merrybent & $<$6.80 & 0.36$\pm$0.14& $<$0.37& $<$0.32& $<$0.19& $<$0.81& $<$2.53& $<$0.14 & 0.22$\pm$0.07 & -& $<$0.06 \\
         & Kamioka & LabC01    & $<$6& $<$0.28& $<$0.3& $<$0.19& $<$4.1& $<$0.77& $<$1& $<$0.091 & 0.34$\pm$0.093& $<$0.061& $<$0.13 \\
210822   & BUGS    & Merrybent & $<$7.83 & 0.62$\pm$0.24& $<$0.49 & 0.38$\pm$0.22& $<$0.34& $<$1.03& $<$2.01& $<$0.17 & 0.35$\pm$0.12 & -& $<$0.11 \\
         & Kamioka & LabC01    & $<$9.4& $<$0.31& $<$0.2& $<$0.33& $<$2.5& $<$0.92& $<$1.4& $<$0.045 & 0.38$\pm$0.1& $<$0.069& $<$0.14 \\
210823   & Kamioka & LabC02    & $<$13& $<$0.38& $<$0.29& $<$0.3& $<$5.2& $<$1.1 & 2.2$\pm$0.84& $<$0.087 & 1.1$\pm$0.15& $<$0.088& $<$0.13 \\
210922   & Kamioka & LabC01    & $<$9.2& $<$0.24& $<$0.46& $<$0.35& $<$2.8& $<$0.82& $<$0.95& $<$0.059 & 0.26$\pm$0.083& $<$0.07& $<$0.14 \\
211006   & Kamioka & LabC01    & $<$9.7& $<$0.22 & 0.39$\pm$0.19& $<$0.24& $<$2.3& $<$0.86& $<$1& $<$0.044& $<$0.14& $<$0.071& $<$0.17 \\
211106   & BUGS    & Belmont   & $<$3.69& $<$0.24 & 0.41$\pm$0.16 & 0.23$\pm$0.12& $<$0.23& $<$0.53& $<$1.23& $<$0.08& $<$0.08 & -& $<$0.04 \\
         & Kamioka & LabC02    & $<$10& $<$0.61& $<$0.43& $<$0.23& $<$2.3& $<$1.3& $<$1.7& $<$0.093& $<$0.25& $<$0.098& $<$0.15 \\
211201   & BUGS    & Merrybent & $<$6.63& $<$0.42 & 0.31$\pm$0.20& $<$0.58& $<$0.21& $<$1.01& $<$1.88& $<$0.11 & 0.19$\pm$0.08& $<$0.11& $<$0.08 \\
         & Kamioka & LabC01    & $<$15 & 0.26$\pm$0.13& $<$0.42& $<$0.19& $<$3.6& $<$0.95& $<$1.1& $<$0.049& $<$0.27& $<$0.072& $<$0.16 \\
211202   & Kamioka & LabC02    & $<$12& $<$0.38& $<$0.39& $<$0.2& $<$2.3& $<$1.2& $<$1.1& $<$0.12 & 0.21$\pm$0.094& $<$0.079& $<$0.12 \\
211204   & Kamioka & LabC01    & $<$8.2 & 0.2$\pm$0.094& $<$0.26& $<$0.38& $<$2.3& $<$0.84& $<$1& $<$0.049& $<$0.18& $<$0.062& $<$0.14 \\
211205   & Kamioka & LabC01    & $<$7& $<$0.3& $<$0.4& $<$0.18& $<$2.3& $<$0.96& $<$0.94& $<$0.064 & 0.29$\pm$0.085& $<$0.086& $<$0.14 \\ \hline

    \end{tabular}
    }
    \caption{\label{tab:GeResult2021}
    Results from HPGe assays of all samples produced in 2021 in units of mBq/kg.
    The upper limit is 95\% CL.
    }
\end{table}
\end{landscape}

\begin{landscape}
\begin{table}[!ht]
    \centering
    \scalebox{0.78}[0.78]{
    \begin{tabular}{r|cc|ccccccccccc}
    \hline

Batch ID & Lab     & detector  & $^{238}$U & $^{226}$Ra & $^{228}$Ra & $^{228}$Th & $^{235}$U & $^{223}$Ra & $^{40}$K & $^{138}$La & $^{176}$Lu & $^{134}$Cs & $^{137}$Cs \\ \hline
220102   & Kamioka & LabC02    & $<$8.2& $<$0.2& $<$0.35& $<$0.19& $<$2& $<$0.84& $<$1.1& $<$0.062& $<$0.15& $<$0.064& $<$0.11 \\
220103   & Kamioka & LabC01    & $<$6.2 & 0.41$\pm$0.12& $<$0.46& $<$0.18& $<$2.5& $<$0.83& $<$1& $<$0.036& $<$0.17& $<$0.068& $<$0.13 \\
220104   & Kamioka & LabC02    & $<$7.1 & 0.35$\pm$0.14& $<$0.32& $<$0.21& $<$2.1& $<$0.92& $<$0.94& $<$0.1 & 0.23$\pm$0.086& $<$0.081& $<$0.11 \\
220201   & Kamioka & LabC01    & $<$8.9 & 0.31$\pm$0.13& $<$0.42& $<$0.25& $<$2.4& $<$0.87& $<$1& $<$0.056& $<$0.31& $<$0.071& $<$0.13 \\
220241   & BUGS    & Belmont   & $<$6.37 & 0.34$\pm$0.11 & 0.22$\pm$0.12& $<$0.30& $<$0.27& $<$0.52& $<$1.28& $<$0.07 & 0.09$\pm$0.04 & -& $<$0.07 \\
         & Kamioka & LabC01    & $<$7.1& $<$0.31& $<$0.5& $<$0.21& $<$2.9& $<$1& $<$1.4& $<$0.041& $<$0.23& $<$0.086& $<$0.14 \\
220242   & LSC     & geOroel   & $<$9& $<$0.21& $<$0.26& $<$0.70& $<$0.36& $<$1.2& $<$1.1& $<$0.2 & 0.69$\pm$0.06& $<$0.08& $<$0.11 \\
         & BUGS    & Belmont   & $<$5.00& $<$0.46& $<$0.28& $<$0.39& $<$0.39& $<$0.79& $<$1.56& $<$0.06 & 0.51$\pm$0.11 & -& $<$0.08 \\
         & Kamioka & LabC02    & $<$7.9& $<$0.49& $<$0.32& $<$0.21& $<$2.3& $<$1.2& $<$1.1& $<$0.07 & 0.53$\pm$0.099& $<$0.074& $<$0.12 \\
220251   & BUGS    & Belmont   & $<$4.59& $<$0.25& $<$0.35 & 0.35$\pm$0.14& $<$0.18& $<$0.63& $<$1.39& $<$0.10& $<$0.13 & -& $<$0.06 \\
         & Kamioka & LabC01    & $<$5.8 & 0.27$\pm$0.1& $<$0.4& $<$0.33& $<$2.1& $<$0.81& $<$1.3& $<$0.038& $<$0.14& $<$0.05& $<$0.21 \\
220351   & LSC     & geAsterix & $<$9& $<$0.19& $<$0.35& $<$0.34& $<$0.47& $<$1.6& $<$1.9& $<$0.16& $<$0.11& $<$0.09& $<$0.10 \\
         & Kamioka & LabC02    & $<$8& $<$0.4& $<$0.41& $<$0.22& $<$2.2& $<$0.96& $<$1.4& $<$0.073& $<$0.13& $<$0.14& $<$0.13 \\
220352   & BUGS    & Belmont   & $<$5.56& $<$0.30& $<$0.48& $<$0.40& $<$0.30& $<$0.53& $<$1.67& $<$0.07& $<$0.12 & -& $<$0.14 \\
         & Kamioka & LabC01    & $<$10& $<$0.31& $<$0.6& $<$0.16& $<$2.5& $<$0.96& $<$1.2& $<$0.047& $<$0.19& $<$0.06& $<$0.15 \\
220353   & LSC     & geAnayet  & $<$49& $<$0.25& $<$0.73& $<$1.3& $<$2.0& $<$2.4& $<$0.78& $<$0.3& $<$0.2& $<$0.14& $<$0.19 \\
         & Kamioka & LabC02    & $<$8.2& $<$0.35& $<$0.42& $<$0.17& $<$2.2& $<$1.1& $<$1.9& $<$0.075& $<$0.16& $<$0.08& $<$0.098 \\
220361   & BUGS    & Belmont   & $<$6.62& $<$0.34& $<$0.38& $<$0.65& $<$0.53& $<$0.81& $<$1.92& $<$0.17& $<$0.18& $<$0.08& $<$0.09 \\
         & Kamioka & LabC01    & $<$10 & 0.48$\pm$0.14& $<$0.34& $<$0.25& $<$2.1& $<$0.91& $<$1.2& $<$0.053 & 0.2$\pm$0.084& $<$0.078& $<$0.15 \\
220371   & LSC     & geOroel   & $<$10& $<$0.26& $<$0.41& $<$0.29& $<$0.45& $<$1.8& $<$1.6& $<$0.2 & 0.15$\pm$0.06& $<$0.11& $<$0.16 \\
         & Kamioka & LabC02    & $<$9.4& $<$0.25& $<$0.43& $<$0.26& $<$2.3& $<$0.98& $<$2.2& $<$0.076& $<$0.16& $<$0.092& $<$0.12 \\
220471   & BUGS    & Belmont   & $<$5.66 & 0.65$\pm$0.21 & 0.77$\pm$0.33 & 0.89$\pm$0.28& $<$0.28& $<$0.78& $<$4.71& $<$0.08& $<$0.21& $<$0.06& $<$0.05 \\
         & Kamioka & LabC01    & $<$6.5 & 0.32$\pm$0.1& $<$0.28& $<$0.26& $<$2.5& $<$0.82& $<$1& $<$0.046& $<$0.29& $<$0.068& $<$0.13 \\
220481   & LSC     & geAsterix & $<$12& $<$0.22& $<$0.41& $<$0.45& $<$0.60& $<$2.4& $<$2.4& $<$0.22 & 0.25$\pm$0.06& $<$0.06& $<$0.1 \\
         & Kamioka & LabC02    & $<$8& $<$0.2& $<$0.6& $<$0.18& $<$2.1& $<$0.91& $<$0.91& $<$0.064& $<$0.14& $<$0.12& $<$0.12 \\
220482   & BUGS    & Merrybent & $<$11.54& $<$0.49 & 0.46$\pm$0.28& $<$0.55& $<$0.25& $<$1.43& $<$2.18& $<$0.20& $<$0.15& $<$0.16& $<$0.17 \\
         & Kamioka & LabC02    & $<$9& $<$0.48& $<$0.36& $<$0.15& $<$3& $<$0.95& $<$2& $<$0.075& $<$0.31& $<$0.071& $<$0.11 \\
220581   & LSC     & geAsterix & $<$12& $<$0.21& $<$0.46& $<$0.49& $<$0.74& $<$2.1& $<$1.6& $<$0.25& $<$0.21& $<$0.08& $<$0.11 \\
         & Kamioka & LabC01    & $<$8.3& $<$0.25& $<$0.24& $<$0.22& $<$2.2& $<$1& $<$1.3& $<$0.06& $<$0.22& $<$0.074& $<$0.17 \\
220582   & BUGS    & Merrybent & $<$16.1& $<$0.67& $<$0.84& $<$0.65& $<$0.36& $<$1.51& $<$3.44& $<$0.22& $<$0.33& $<$0.11& $<$0.12 \\
         & Kamioka & LabC02    & $<$9.1& $<$0.43& $<$0.6& $<$0.23& $<$2.3& $<$1.3& $<$2.7& $<$0.073& $<$0.26& $<$0.087& $<$0.11 \\
220691   & BUGS    & Merrybent & $<$13.4& $<$0.93& $<$1.04& $<$0.73& $<$0.33& $<$1.82& $<$2.90& $<$0.23& $<$0.28& $<$0.14& $<$0.22 \\
         & Kamioka & LabC01    & $<$6.5& $<$0.33& $<$0.22& $<$0.17& $<$2.1& $<$0.88& $<$1.4& $<$0.049& $<$0.29& $<$0.077& $<$0.13 \\
220603   & Kamioka & LabC02    & $<$9.3& $<$0.44& $<$0.47& $<$0.24& $<$2.4& $<$1.3& $<$2& $<$0.055& $<$0.29& $<$0.083& $<$0.13 \\ \hline

    \end{tabular}
    }
    \caption{\label{tab:GeResult2022}
    Results from HPGe assays of all samples produced in 2022 in units of mBq/kg.
    The upper limit is 95\% CL.
    }
\end{table}
\end{landscape}





\end{document}